\documentclass[12pt]{iopart}
\usepackage{iopams}  
\usepackage{graphicx}
\expandafter\let\csname equation*\endcsname\relax
\expandafter\let\csname endequation*\endcsname\relax
\usepackage{amsmath}
\usepackage{mathabx}
\usepackage{cite}
\usepackage{lscape}
\usepackage{varwidth}
\usepackage{textcomp}
\usepackage{xcolor,soul}
\usepackage{color,soul}
\usepackage{algorithm}
\usepackage{algpseudocode}
\usepackage{bm}
\usepackage{esvect}
\usepackage{accents}
\usepackage{caption}
\usepackage{float}

\begin{document}

\title[MICROSCOPE satellite]{MICROSCOPE Satellite and its Drag-Free and Attitude Control System}

\author{Alain Robert$^1$, Valerio Cipolla$^1$, Pascal Prieur$^1$, Pierre Touboul$^2$, Gilles M\'etris$^3$, Manuel Rodrigues$^2$, Yves Andr\'e$^1$, Joel Berg\'e$^2$, Damien Boulanger$^2$, Ratana Chhun$^2$, Bruno Christophe$^2$, Pierre-Yves Guidotti$^1$ \footnote{Current address: AIRBUS Defence and Space, F-31402 Toulouse, France}, Emilie Hardy$^2$, Vincent Lebat$^2$, Thomas Lienart$^1$, Fran\c{c}oise Liorzou$^2$, Benjamin Pouilloux$^1$ \footnote{Current address: KINEIS, F-31520 Ramonville Saint-Agne, France}}

\address{$^1$ CNES, 18 avenue E Belin, F-31401 Toulouse, France}
\address{$^2$ ONERA, Universit\'e Paris Saclay, F-92322 Ch\^atillon, France}
\address{$^3$ Universit\'e C\^ote d{'}Azur, Observatoire de la C\^ote d'Azur, CNRS, IRD, G\'eoazur, 250 avenue Albert Einstein, F-06560 Valbonne, France}

\ead{alainjm.robert@cnes.fr, pierre.touboul@onera.fr, gilles.metris@oca.eu, manuel.rodrigues@onera.fr}
\vspace{10pt}
\begin{indented}
\item[] Dec. 2020
\end{indented}

\begin{abstract}
This paper focus on the description of the design and performance of the MICROSCOPE satellite and its Drag-Free and Attitude Control System (DFACS). The satellite is derived from CNES' Myriade platform family, albeit with significant upgrades dictated by the unprecedented MICROSCOPE's mission requirements. The 300kg drag-free microsatellite has completed its 2-year flight with higher-than-expected performances. Its passive thermal concept allowed for variations smaller than 1\,$\mu$K at the measurement frequency $f_{\rm{EP}}$. The propulsion system provided a 6 axis continuous and very low noise thrust from zero to some hundreds of micronewtons.  
Finally, the performance of its DFACS (aimed at compensating the disturbing forces and torques applied to the satellite) is the finest ever achieved in low Earth orbit, with residual accelerations along the three axes are lower than $10^{-12} m/s^2$ at $f_{\rm{EP}}$ over 8 days. 

\end{abstract}


\submitto{\CQG}
%
%
%

\section{Introduction}
The weak equivalence principle (WEP) states that all bodies should fall at the same rate in a given gravitational field, independently of their mass or composition. MICROSCOPE is a CNES-ESA-ONERA-CNRS-OCA-DLR-ZARM fundamental physics mission dedicated to the test of the WEP in space. The MICROSCOPE satellite aimed at testing its validity at the $10^{-15}$ precision level by measuring the force required to maintain two test masses (one made of titanium and the other of platinum alloys) exactly in the same orbit. 
The microsatellite was launched in 2016 into an altitude of 710 km dawn-dusk sun-synchronous orbit (6PM at the ascending node).  The scientific mission requires extremely accurate control of the linear and angular accelerations of the satellite. In science mission mode, the propulsion subsystem continuously overcomes the non-gravitational forces and torques (air drag, solar pressure, magnetic torques, etc.) in such a way that the satellite follows the test masses in their pure gravitational motion. The satellite is also spun about the normal axis to the orbital plane in order to increase the modulation frequency of the Earth's gravity field and thus the frequency of the potential WEP signal, $f_{\rm{EP}}$.

The paper aims at describing the satellite and its main subsystems entering in the mission performance budget such as the DFACS, the propulsion subsystem and the precise orbit determination. The mission requirements in \cite {rodriguescqg1} have been distributed on these subsystem and are described here. The paper concludes with the performance observed in orbit.


\section {Satellite description} \label{sect:sat}
\subsection {Overview} \label{sect:satgen}

The MICROSCOPE spacecraft \cite{Cipolla17, Cipolla11} has been developed within the framework of the Myriade micro-satellite product line. This line is dedicated to performing scientific or propaedeutic missions with reduced development schedule and costs targeting payload in the class of 60 kg / 60 W. Because of MICROSCOPE's challenging goals, the platform was specifically adapted in size but still with the line equipment.
The architecture of the standard Myriade satellite is based on a platform with generic functional chains and on a mission-customized payload usually located on the top of the platform structure.
In the case of the MICROSCOPE satellite, the design was constrainted by (i) the needs to control the acceleration of the satellite along its six degrees of freedom; (ii) the implementation of the payload as main attitude and orbit control system (AOCS) sensor; (iii) the distinctiveness of payload interface (I/F) and characteristics and (iv) the minimisation of the mass motions.    

Since MICROSCOPE is more demanding in terms of performance and interface constraints, the Myriade generic platform could have been adapted, and numerous elements of the microsatellite line were reused as the Ground Support Equipment, the structural concepts and the integration and validation principles.
In order to minimize risks and additional costs, particular developments were limited to the functional chain playing a key role in the mission.
In addition to the payload described in \cite{liorzoucqg2}, the satellite subsystem are the satellite structure, the propulsion, the thermal control, the power supply, the command control, the telemetry/telecommand (TM/TC), the AOCS/DFACS, the navigation system (GNSS, i.e GPS) and the de-orbitation system.

\begin{figure} [h]
\begin{center}
\includegraphics[width=0.7\textwidth]{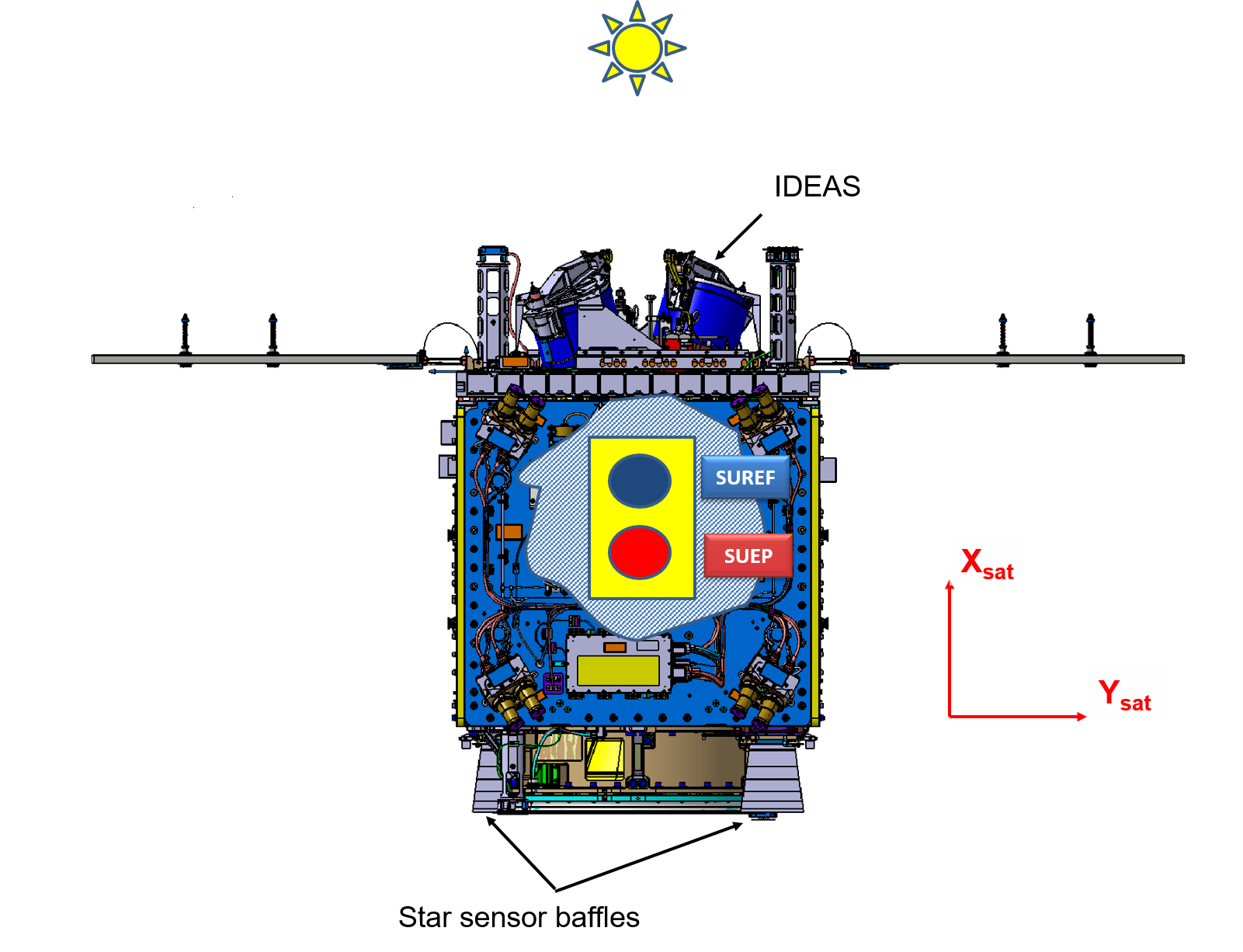}
\includegraphics[width=0.7\textwidth]{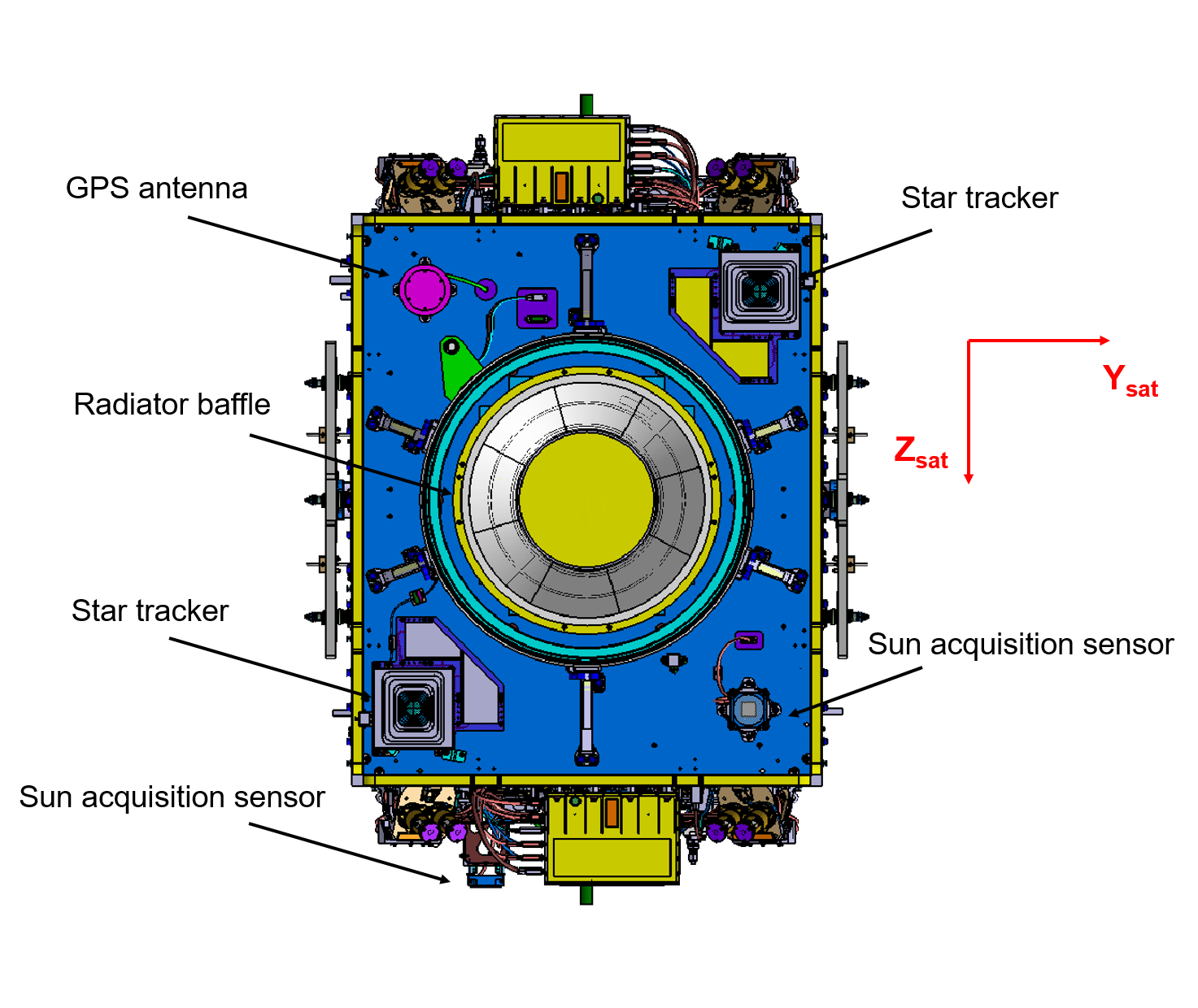}
\caption{Upper panel: satellite reference frame with two disks illustrating the cylinder section of the two sensor units with the two units containing the deorbitation wings. Lower panel: the bottom part of the satellite with the solar panels closed in launch position, the radiator baffle in the centre with the two star trackers on opposite sides.}
\label{fig_sat}       
\end{center}
\end{figure}

The spacecraft (s/c) flight configuration (i.e. with solar panel deployed) is around 1.36 m long (along $X_{\rm{sat}}$ axis), 2.78 m wide (along $Y_{\rm{sat}}$ axis) and 1.28 m deep (along $Z_{\rm{sat}}$ axis) : for cross sections of 1.52\,m$^2$ normal to $Y_{\rm{sat}}$ and 1.12\,m$^2$ normal to $Y_{\rm{sat}}$).
Fig. \ref{fig_sat} gives an overview of the satellite.

\subsubsection {Satellite reference frame}

The reference frame (see Fig. \ref{fig_sat}) is defined as follows: 
\begin{itemize}
\item The $X_{\rm{sat}}$ is perpendicular to Launcher I/F ring, while in orbit it corresponds to the spin axis.
\item The $Y_{\rm{sat}}$ is defined by the direction of the two solar arrays (in stowed configuration).
\item The $Z_{\rm{sat}}$ completes the triad and is aligned to the payload sensitive axis (cylinder axis).
\end{itemize}

\subsubsection {Mass, centre of mass and inertia budget}
The total mass at launch was 301.6 kg including 16.5 kg of propellant (see section \ref{sect_cgps}). 
The center of gravity of the satellite was balanced to the accuracy of the test-bench of $\pm$1.4\,mm. Exploitation of inflight data leads to a better estimation of 0.4\,mm.
To minimize the satellite unbalance parasitic forces when spinning out of the spacecraft center of mass, inertia moments dissymetries have been measured to $\pm$4 kg\,m$^2$ accuracy. 

\subsection {Structure} \label{sect:structure}

\subsubsection {Payload module}
\begin{figure}
\begin{center}
\includegraphics[width=0.4\textwidth]{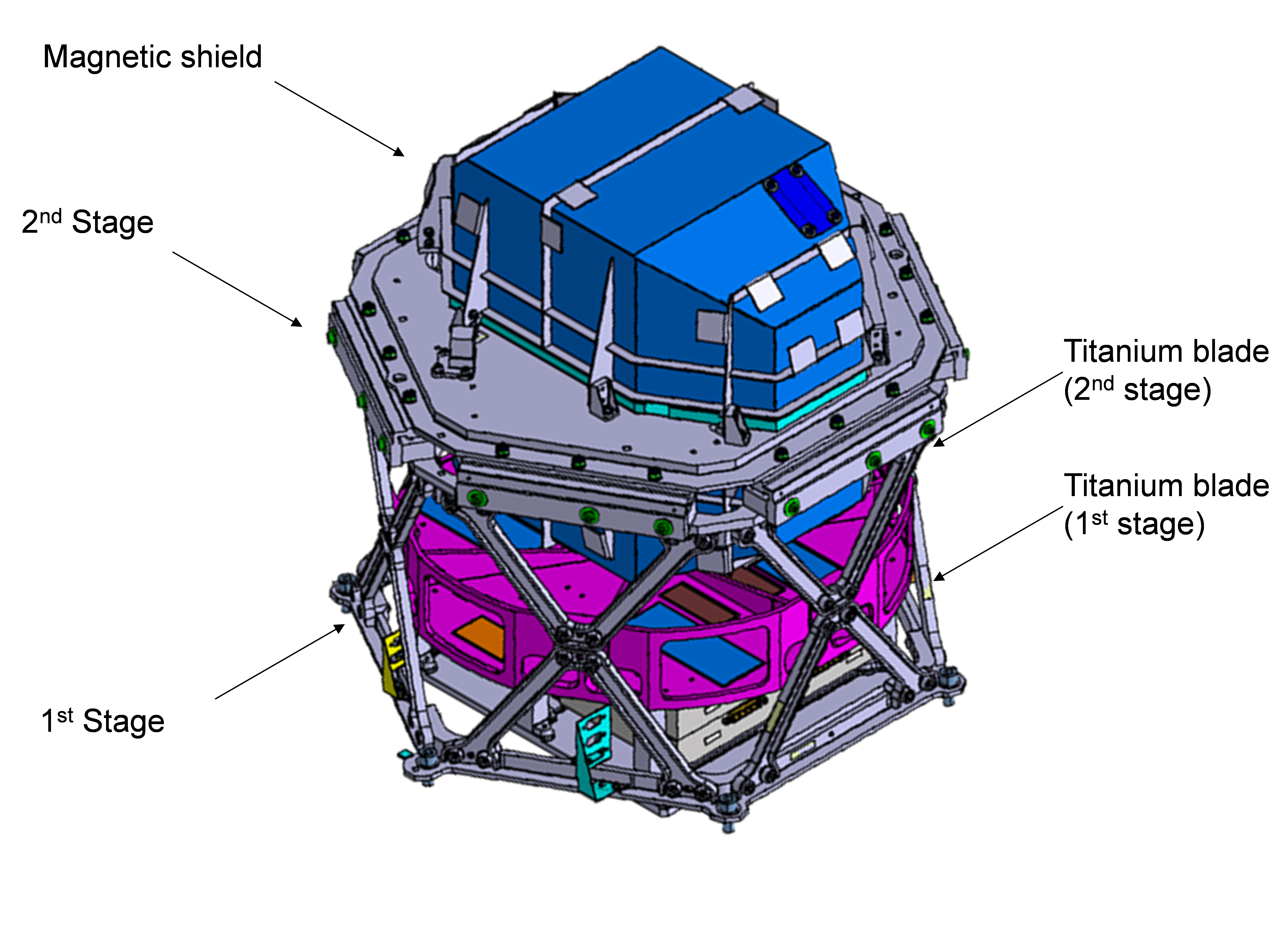}
\includegraphics[width=0.4\textwidth]{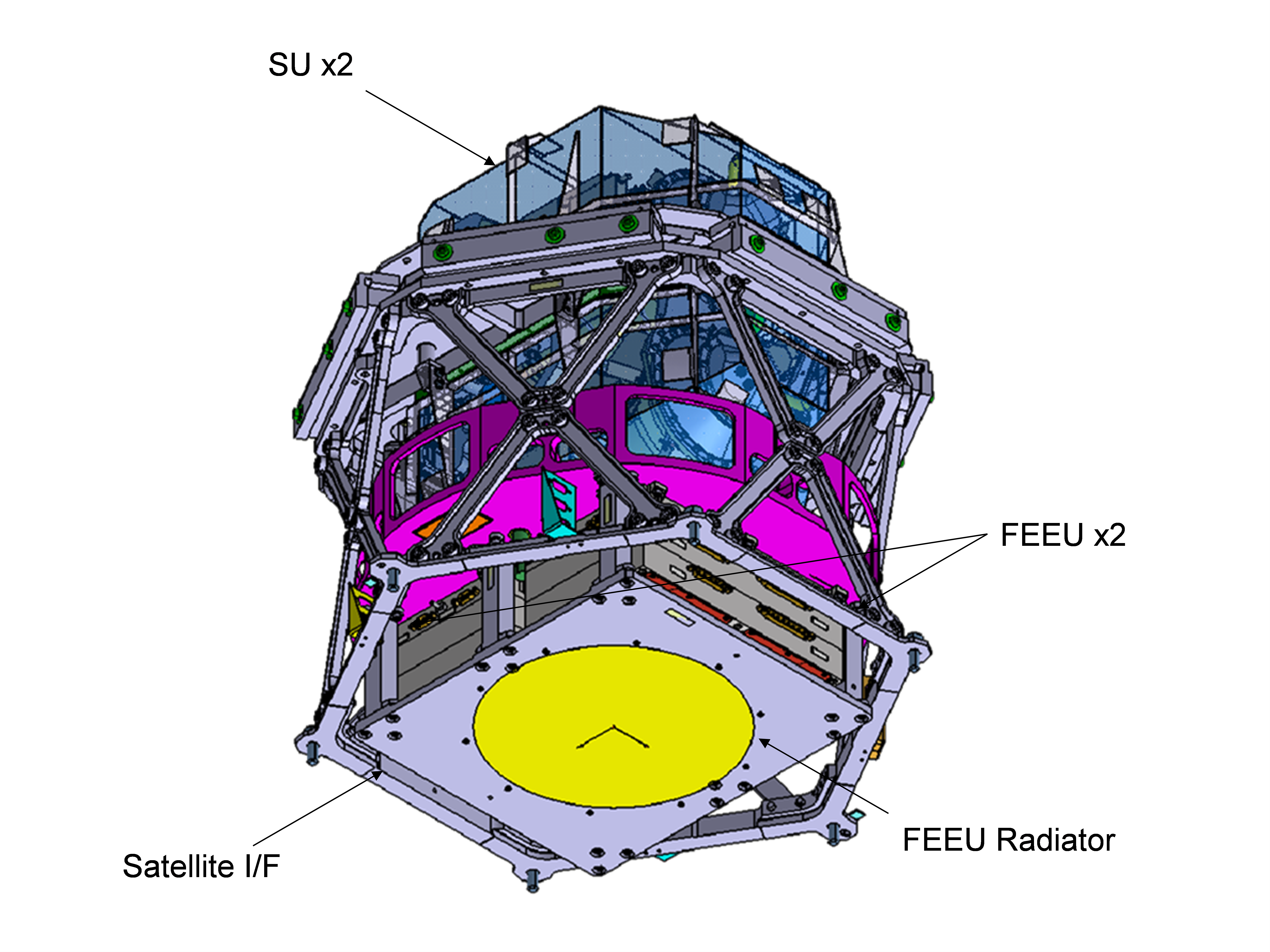}
\caption{Payload module design}
\label{fig_bcu}       
\end{center}
\end{figure}

The scientific instrument, called T-SAGE for "Twin - Spaced Accelerometer for Gravity Experiment", is accommodated inside the payload module (Fig. \ref{fig_bcu}). 
It comprises two sensor units (SU) and two associated Front-End Electronics Units (FEEU); see Ref. \cite{liorzoucqg2}. Each SU is a double electrostatic accelerometer composed of two cylindrical and concentric test masses. One SU comprises two test masses of same material for the check of the mission, while the other SU comprises two test masses made of different material for the principle equivalence test. The accelerometer measures the difference of accelerations \cite{rodriguescqg1} needed to be applied on the two test-bodies to keep identical motion while submitted to the same gravity. 
The payload module is mounted on the panel opposite the Sun in mission mode, for thermal stability reasons. It is located as close as possible to the center of the satellite in order to minimize the external perturbations. The satellite integration ensures that the test-mass centers of gravity are at less than 5\,mm from the satellite centre along $Y_{\rm{sat}}$ and $Z_{\rm{sat}}$.

The payload module was specified to provide a thermal stability of 1\,mK at $f_{\rm{EP}}$ at the SU interface, and of 10\,mK at $f_{\rm{EP}}$ at the FEEU interface \cite{hardycqg6}.
The payload module is also designed to guarantee good mechanical alignment and stability with respect to the star sensor interface.  Finally, the payload module is covered with a magnetic shield in order to limit the effect of magnetic field disturbances in the SU.
It is designed as a two-stage system. The first stage accommodates both FEEU and its radiator: it is fixed to the platform through an interface ring by six titanium blades which guarantee the thermal decoupling from the rest of the satellite. The second stage accommodates the SU and its magnetic shield: it is fixed to the first stage by six titanium blades which guarantee the thermal decoupling.
Both stages are insulated by a Multi Layer Insulator (MLI) to reduce radiative thermal exchange between the different stages with respect to the platform.

\subsubsection {Satellite structure.}
The MICROSCOPE structure has been designed and manufactured based on the same principle and material of the Myriade generic bus but with different dimensions and equipment accommodations.

The compatibility of the structure with the auxiliary payload launching system, Soyouz ASAP-S, has been kept as design driver. The main structure is composed of six walls made with an honeycomb core and Aluminum alloy skins, linked by Aluminum alloy corners. 

One wall accommodates the payload module, the star tracker optical heads and the launcher interface structure and adaptor. Another one accommodates the IDEAS deorbiting system (see section \ref{sect_ideas}). Two walls accommodate the Cold Gas Propulsion System (CGPS - section \ref{sect_cgps}). The remaining walls are dedicated to classic Myriade platform functional chain (Power, avionics, telemetry/telecommand, AOCS) and GNSS.
The layout of the equipment has been optimized in order to balance the mass with respect to rotation axis and to minimize the inertia cross-products.

\begin{figure}
\begin{center}
\includegraphics[width=0.42\textwidth]{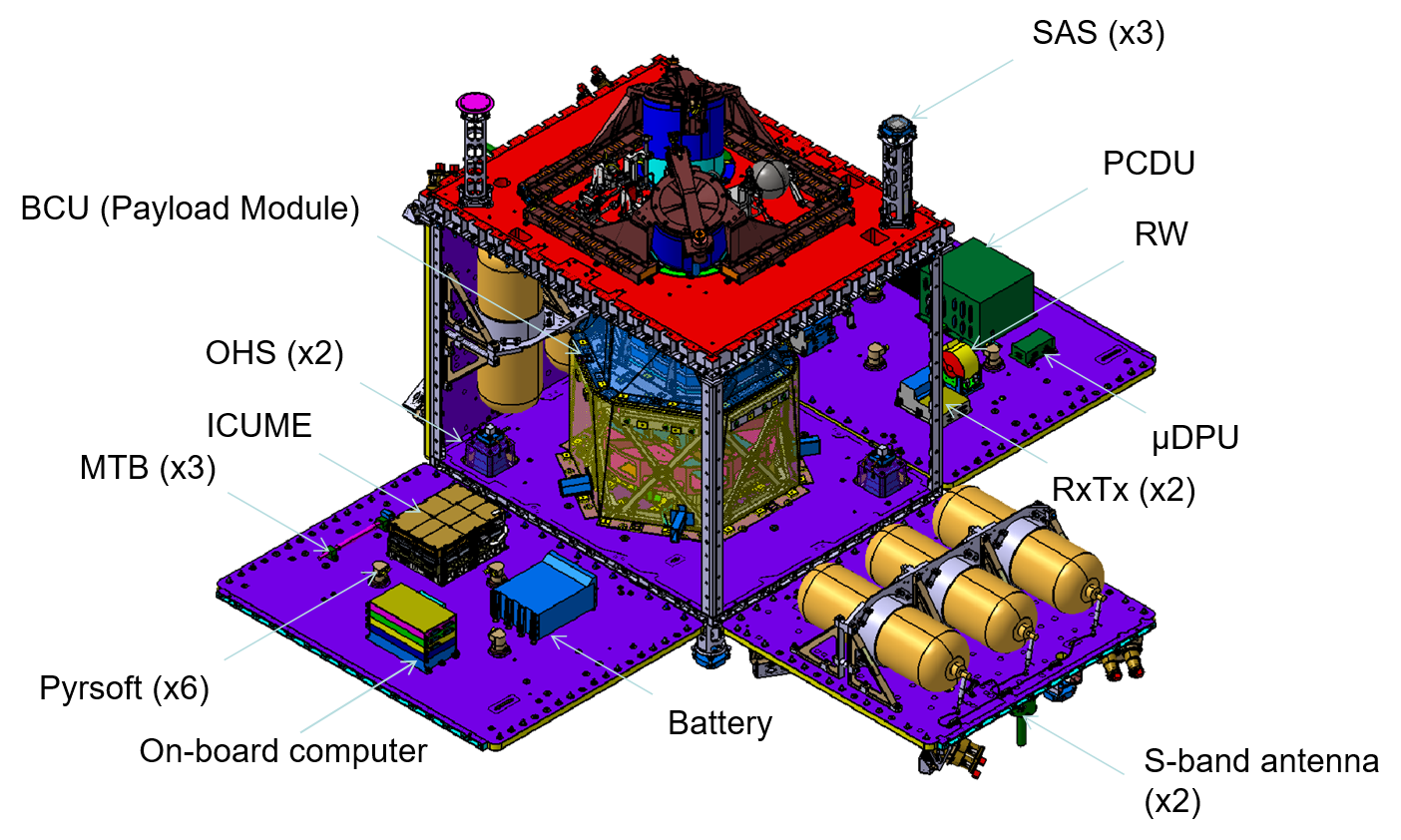}
\includegraphics[width=0.42\textwidth]{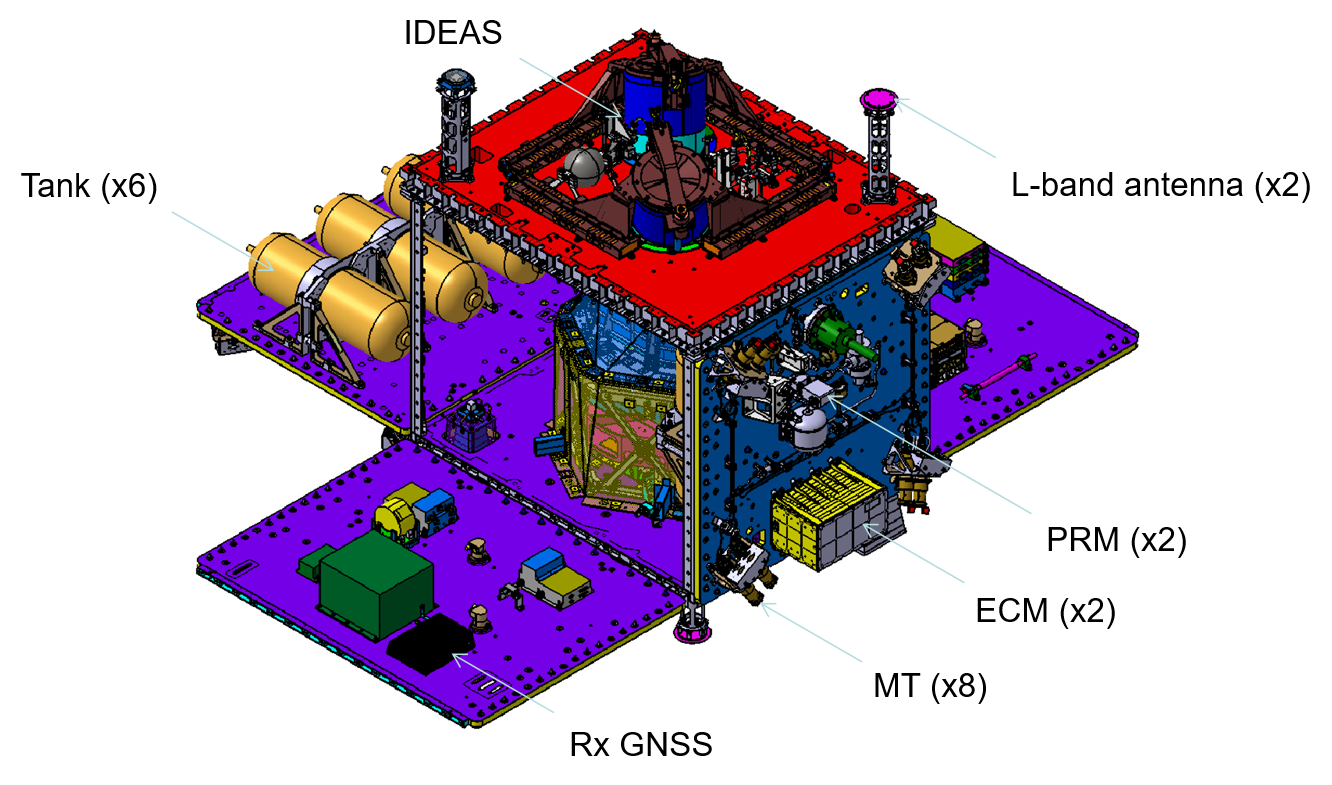}
\caption{Payload module and Myriade equipment accommodation}
\label{fig_equi}       
\end{center}
\end{figure}

This modularity allowed us to optimize the acceptance and integration test (AIT) schedule by performing several integrations in parallel (payload module, CGPS, IDEAS and platform).

\subsection {Thermal control} \label{sect:thermal}
From a thermal point of view, the satellite can be decomposed in several thermal cavities, each with its own idependant thermal control : payload module stage, star tracker optical heads, CGPS tanks, CGPS electronic module, satellite structure.

In order to minimize the power budget and thermal disturbances, the thermal control design is based mainly on passive principles using MLI and dedicated radiators.
Heaters are used only in safe mode and transition mode to keep the temperature of equipment inside their operative or non-operative range; in science mission mode, heater activation is forbidden in order to protect the payload against electromagnetic perturbations.
The platform radiators are the main sources of thermal disturbance to the instrument; for this reason, they are accommodated symmetrically and their surfaces have been defined in order to be as nearly as possible the same. However, the radiator for the FEEU thermal control is located at the launcher interface and cannot be balanced by a symetrical one. In order to avoid any entrance at $f_{\rm{EP}}$ from Earth's abeldo through this radiator into the payload module, the radiator is protected by a dedicated baffle.
The performances of the thermal control are significantly better than the requirements; in-flight results are summarized in Table \ref{tab_therm}, see Ref. \cite{torresi, hardycqg6} for more details.

\begin{table}
\caption{\label{tab_therm} In-flight observed thermal variations at payload level.}
\begin{indented}
\item[]\begin{tabular}{@{}lll}
\br
Location: s/c mode & Requirement (0-peak) & In flight results  \\
\mr
FEEU: inertial session  & $<10$ mK at $f_{\rm{EP}}$ & $<4$ mK \\
\mr
FEEU: rotating session & $<3$ mK at $f_{\rm{EP}}$ & $<0.4$ mK for the S/C spin rate 2 \\
 & & $<0.08$ mK for the S/C spin rate 3\\
\mr
SU: inertial session & $<1$ mK at $f_{\rm{EP}}$ & $<0.5$ mK   \\
\mr
SU: rotating session & $<1$ mK at $f_{\rm{EP}}$ & $<0.8$ $\mu$K for the S/C spin rate 2  \\
  &   & $<0.2$ $\mu$K for the S/C rate 3  \\
\br
\end{tabular}
\end{indented}
\end{table}

\subsection {Avionics and Command/Control}
\begin{figure}
\begin{center}
\includegraphics[width=0.75\textwidth]{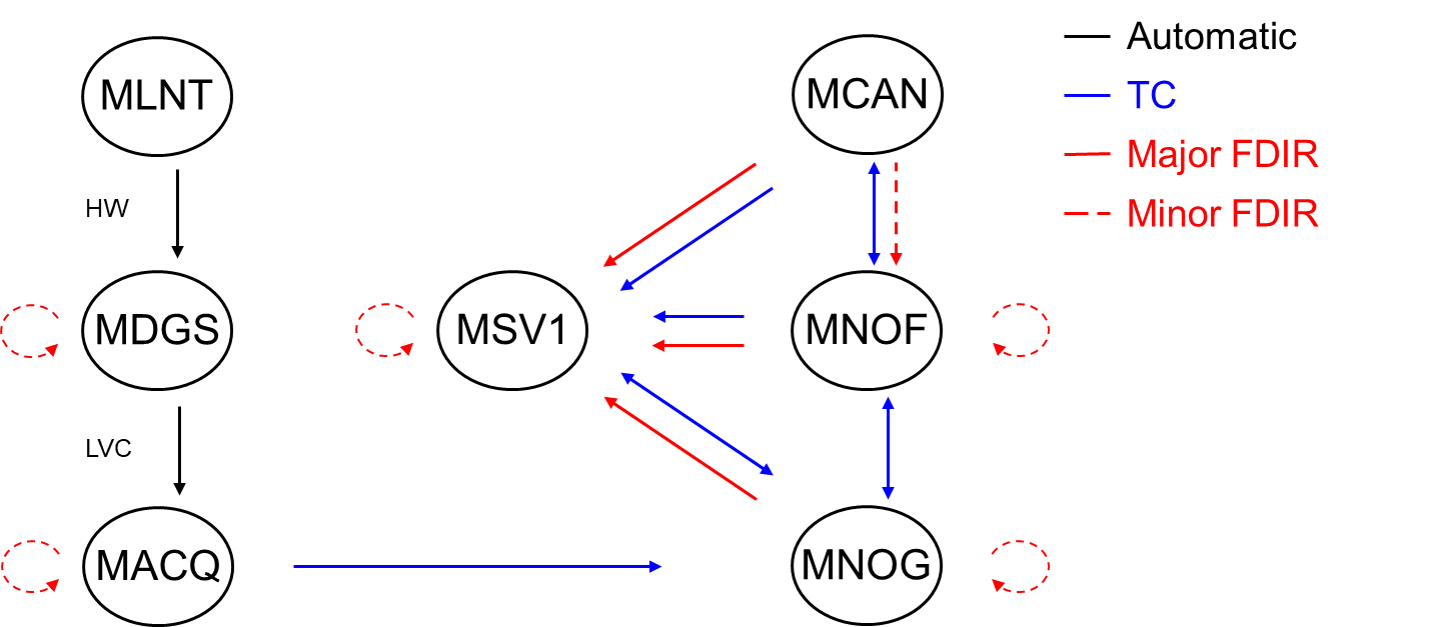}
\caption{Main satellite modes and transitions (HW: hardware, SW: software).}
\label{fig_mode}       
\end{center}
\end{figure}

MICROSCOPE reuses the same on-board computer as the Myriade product line based on a T805 $\mu$processor and commercial off-the-shelf (COTS) components. It provides 1 GB of memory for housekeeping (HK) and science telemetry (TM) and up to 5 MIPS through an S-Band link of 625kbit/s.
The satellite modes and transitions are shown in Fig. \ref{fig_mode} :
\begin {itemize}
\item {\bf MNLT} corresponds to Launch Mode: the satellite is off and only the separation detection circuit is powered.
\item {\bf MDGS} corresponds to solar array deployment mode: the satellite is switched on after the separation detection and the solar array is deployed after a countdown.
\item {\bf MACQ} corresponds to first acquisition mode: the satellite automatically points its solar array toward the Sun and spins slowly around Sun direction to maximize the available power and stabilize the temperature.
\item {\bf MNOG} corresponds to a transition mode with a coarse pointing; this mode is used for commissioning all the equipment and as withdrawal mode during mission deterministic interruptions (eclipse season, moon transition, etc.).
\item {\bf MNOF} is inherited from the Myriade product line and with the use of CGPS as main actuator; it corresponds to a transition mode with a fine pointing that can also be used  in the collision-avoiding operations.
\item {\bf MCAN} is a new mode that corresponds to the science mission mode; it is detailed in section \ref{sect_guid}.
\item {\bf MSV1} corresponds to a safe mode; its characteristics are the same as MACQ.
\end {itemize}
The MNLT, MDGS, MACQ and MNOG are generic Myriade modes which reuse the product line onboard software.

\subsection {Power supply}
The power supply chain reuses the same equipment as the Myriade product line.
The solar array is composed of two panels; it uses the same cells disposed in a specific layout minimizing its magnetic momentum.
Total surface is $0.84$\,m$^2$ and maximum power is around 240\,W.

The solar panels are stowed during launch; each panel is released after the separation using 3 pyro-nut mechanisms and deployed by one Carpentier blade. Because the selected orbit and satellite attitude guarantee a good sunlight ratio, no solar array drive mechanism has been used.

the s/c battery is a standard Myriade product line equipment made with Li-Ion cells; it provides a maximum energy of 390\,Wh and a maximum capacity of 13.5\,Ah.

The Power Conditioning and Distribution Unit (PCDU) ensures the regulation of the power generated by the solar array (up to 8\,A). It distributes regulated buses (-15\,V, +5\,V, +12\,V, +15\,V and +20\,V) and non-regulated buses (between 22\,V and 37\,V). The payload uses the non-regulated buses. 
The PCDU also provides the battery regulation, the generation of magneto-torque commands and the distribution of pyro commands (up to 12 power lines).

\subsection {Attitude and Orbit Control System (AOCS)} \label{sect_aocs}
The AOCS is used in low level modes such as the launch and early orbit phase sequence, the transition modes and the safe mode. In science mission modes, the AOCS is replaced by the DFACS (see section \ref{sect_guid}).

The AOCS manages 3 modes:
\begin{itemize}
\item MAS corresponds to the s/c safe mode; $X_{\rm{sat}}$-axis is pointed toward the Sun with an accuracy of $20^o$ and the satellite is spun about $X_{\rm{sat}}$ at $0.25^o$/s ($4.36\times 10^{-3}$rad/s).
\item MGT3 corresponds to coarse pointing transition mode; $X_{\rm{sat}}$-axis is pointed perpendicular to the orbit with a $10^o$ tilt (conical) and the satellite is spun about $X_{\rm{sat}}$ with an angular speed of 3 times the orbital period.
\item MSP: corresponds to fine pointing transition mode: $X_{\rm{sat}}$-axis is pointed perpendicular to the orbit with an inertial attitude pointing.
\end {itemize}

AOCS mainly uses the sensors and the actuators of the Myriade line equipment:
\begin{itemize}
\item Three Sun Acquisition Sensors (SAS) with a hemispheric field of view make it possible to determine Sun direction; two are accommodated along the same axis ($X_{\rm{sat}}$-axis) with opposite direction, the third along a perpendicular direction ($+Z_{\rm{sat}}$-axis)  
\item One Magnetometer (MAG) performs the measurement of the 3 components of the magnetic field in the range of $\pm 60\,\mu$T.
\item One Reaction wheel provides a momentum of 0.12\,N\,m\,s and a maximum torque of 5\,mN\,m.
\item Three Magneto Torque Bar (MTB) along three degrees of freedom, capable of generating a magnetic moment of $12$\,A\,m$^2$.
\item A star tracker assembly with two optical heads co-aligned and directed toward $-X_{\rm{sat}}$ direction (anti-solar).
\end {itemize}

The star tracker performances at 3$\sigma$ are:
\begin{itemize}
\item RMS noise of $250\,\mu$rad along line of sight and $30\,\mu$rad perpendicular;
\item Bias of $200\,\mu$rad along line of sight and $50\,\mu$rad perpendicular.
\end {itemize}

Table \ref{tab_aocs} summarizes the correspondence between satellite modes, AOCS modes and equipment used.

\begin{table}
\caption{\label{tab_aocs} Link between satellite modes and AOCS modes.}
\begin{indented}
\item[]\begin{tabular}{@{}llll}
\br
S/C & AOCS & S/C & main active    \\
mode & mode & control & equipments \\
\br
MNLT  & none \\
\mr
MDGS  & none\\
\mr
MACQ & MAS & Sun coarse pointing & SAS, MAG\\
MSV1 & & Slow rate spin around $X_{\rm{sat}}$ & reaction wheel, MTB \\
\mr
MNOG & MGT3 & $X_{\rm{sat}}$ normal to orbit with $10^o$ of tilt & MAG\\
& & S/C spun around $X_{\rm{sat}}$  & reaction wheel, MTB \\
& & at 3 times orbital period &  \\
\mr
MNOF & MSP & 3 axis control & star tracker, CGPS \\
\br
\end{tabular}
\end{indented}
\end{table}

\subsection {Global Navigation Satellite System - GNSS} \label{sect:gnss}

G-SPHERE-S \cite{junique} is a new spatial single-frequency GPS receiver manufactured by SYRLINKS, and is the outcome of a CNES R\&D program aimed at the design of a low cost GNSS software receiver based on COTS components.
The equipment of weight 1\,kg and of a consumption of 4\,W is composed of 3 modules:
\begin{itemize}
\item one DC/DC power module;
\item one numerical module based on a commercial Digital Signal Processor which provides I/F with the on-board computer, radio frequency signal conversion and processing and clock generation;
\item one radio frequency module filtering and conditioning the signal.
MICROSCOPE was the first flight opportunity for this new GNSS.  
\end{itemize}

The receiver software is highly configurable and the performance has already been improved using CNES orbit determination team analysis during ground tests and the commissioning phase. The MICROSCOPE satellite rotates around $X_{\rm{sat}}$, which is also the cross-track axis, and the two antennas are placed on $X_{\rm{sat}}$ opposite faces (see Fig. \ref{fig_sat}) in order to collect GPS signals transmitted to the receiver through an analog coupler.
GNSS performances are described in more detail in section \ref{sect_gps}.

\subsection {Innovative DEorbiting Aerobrake System - IDEAS.} \label{sect_ideas}
Because of the low ballistic coefficient of the satellite, the time needed to de-orbit exceeds the limitation of 25 years enforced by French Space Law. This led to the development of a deorbiting system. This system has little impact on mass, volume and power. The selected passive de-orbiting system called IDEAS (Innovative DEorbiting Aerobrake System) fulfills the requirements and has much less acceleration perturbation than chemical propulsion with tank sloshing.
IDEAS is composed of two parts:
\begin{itemize}
\item two identical wings, each including two sails of 4.54\,m length and 0.39\,m width and a Gossamer mast ensuring its deployment,
\item a vessel of 290bar which inflates the sail masts with Nitrogen.
\end{itemize}

\begin{figure}
\begin{center}
\includegraphics[width=0.4\textwidth]{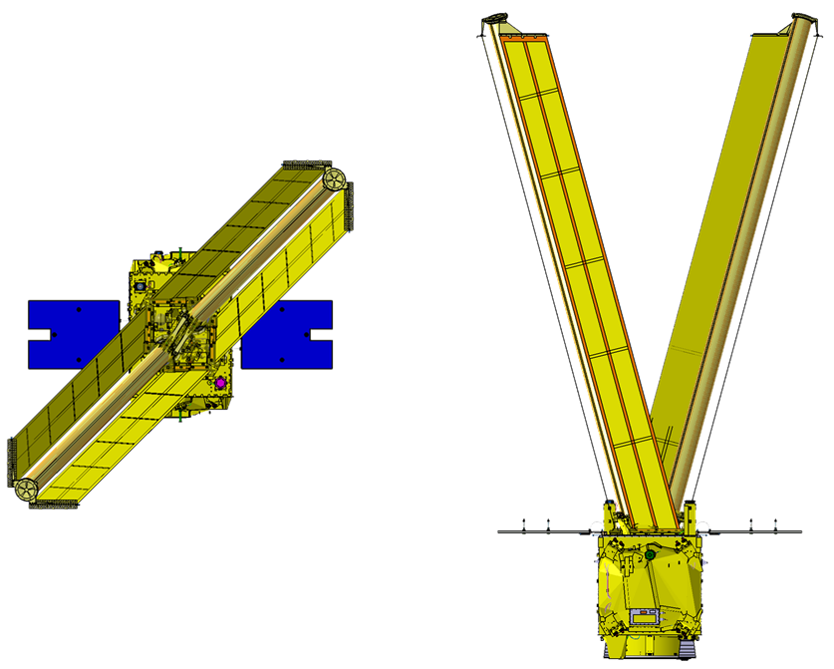}
\includegraphics[width=0.5\textwidth]{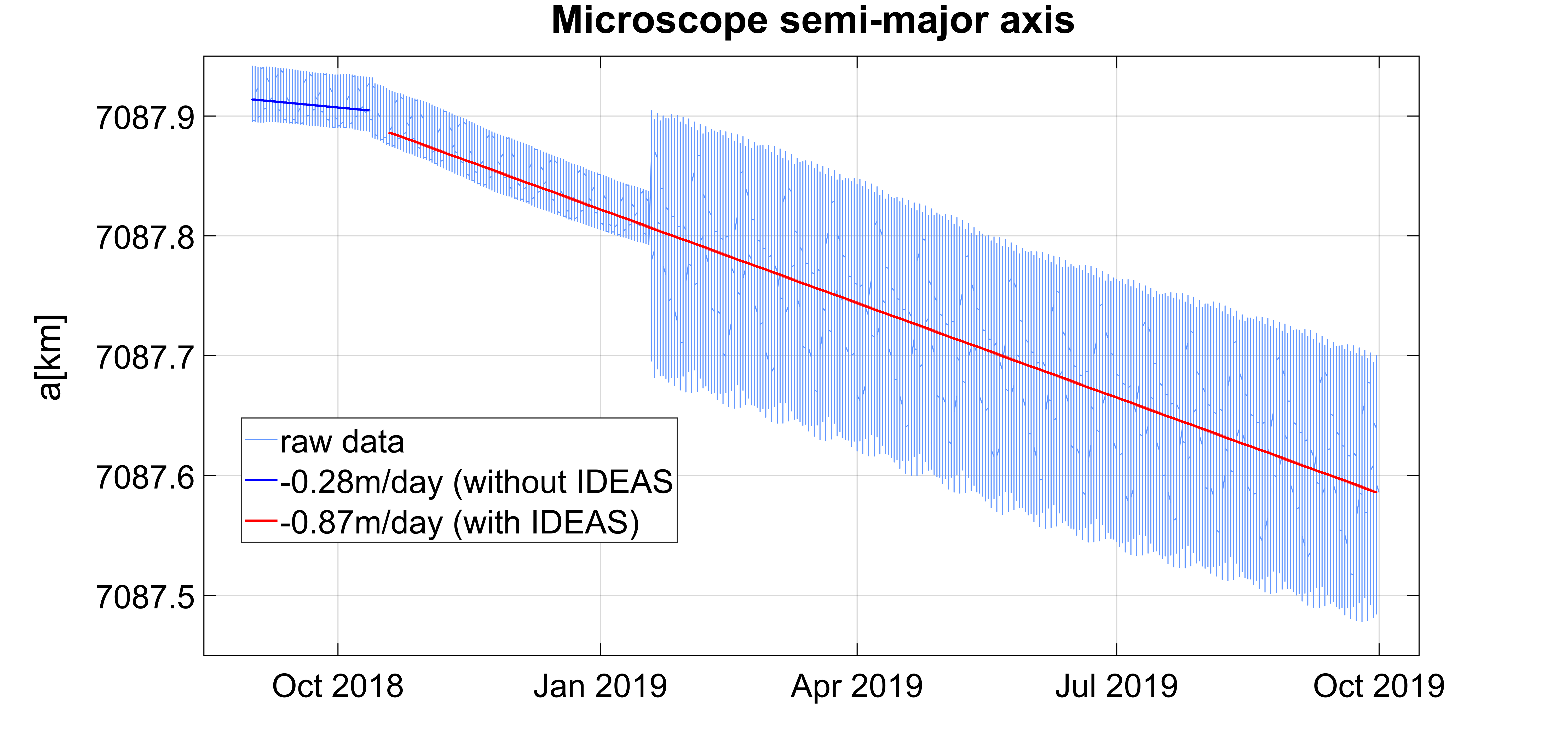}
\caption{Left: satellite after IDEAS deployment. Right: MICROSCOPE semi-major axis evolution a[km] before and after IDEAS deployment.}
\label{fig_sail}       
\end{center}
\end{figure}

Its operating principle is to deploy at the end of the mission braking surfaces (i.e wings) that increase the atmospheric drag of the satellite, accelerating the natural reduction of the orbit. The geometry of the wings has been optimized in order to maximize the ratio between the deployed surface and the mass of de-orbiting system. In addition the design minimizes, after their deployment, the difference between the minimum and the maximum drag surfaces w.r.t. the attitude of the satellite (Fig. \ref{fig_sail}). The drag area of the satellite increases from $2.09$\,m$^2$ to $5.44$\,m$^2$. In Fig. \ref{fig_sail}, the efficiency of the drag is shown throut satellite tracking from Earth. The decrease rate of the orbit semi-major axis goes from $-0.28$\,m\,$/$day to $-0.87$\,m\,$/$day after release .

\section {DFACS design and needs} \label{sect_guid}

\subsection{Drag-Free and Attitude Control System (DFACS) motivations and challenges} \label{sect_chal}

The ``drag-free'' mode stands for a sattelite mode where it is accelerated to compensate the air drag and all  other forces applied on the satellite (Sun radiation and Earth radiation mainly). There are basically two ways of performing ``drag-free'' in space. The first way consists in keeping a test-mass freely floating in a cage as inertial reference; the satellite (or the cage) follows the motion of the test-mass.  The satellite propulsion system applies the needed thrusts to overcome the perturbations so that the test-mass remains centred. The second way consists in measuring the acceleration of the test-mass with respect to the satellite and applying the needed thrusts to nullify the output of the accelerometer. As these accelerations are determined by the forces applied to the satellite, these latter are totally compensated by the propulsion system. In the second case, the inertial mass is the test-mass of an accelerometer. The test-mass is not free but controlled about its six degrees of freedom. 

In the particular case of MICROSCOPE, in order to minimize the thrust consumption, the accelerometer linear measurement bias is estimated and subtracted from the DFACS command. The DFACS is able to control the satellite on a linear combination of the four test masses' six degrees of freedom. 

The DFACS specifications come from the measurement equation detailed in \cite {rodriguescqg1}. For the purpose of this paper, the difference of the applied acceleration to a perfect concentric pair of test masses can be simply expressed by:
\begin{equation}
\vv{\Gamma_d}=\vv{\Gamma_1} - \vv{\Gamma_2}= \delta(2,1) \vv{g},
\label{eq_meas}
\end{equation}
where $\vv{\Gamma_i}$ is the acceleration applied on the test mass $i$ ($i=1$ for inner test mass and $i=2$ for the outer test mass); $\vv{g}$ is the Earth's gravitation field; $\delta(2,1)=\delta$ is a good approximation \cite{touboul19, rodriguescqg1} of the E\"otv\"os parameter of material 2 versus 1 defined by:
\begin{equation} \label{eq_eotvos}
\delta(2,1) =  \frac{m_{g2}}{m_{i2}} - \frac{m_{g1}}{m_{i1}}.
\end{equation}
The actual instrument may present very small differences in the scale factor of the 2 test-masses (respectively $K_1$ and $K_2$).
And thus, the measured signal $\vv{\Gamma_{md}}$ is expressed by $\vv{\Gamma_{md}}=K_1\vv{\Gamma_1} - K_2\vv{\Gamma_2}$.
The measurement equation is also expressed in terms of  common-mode acceleration (i.e. the mean applied acceleration to both concentric test masses) and differential-mode acceleration (the difference of applied acceleration):
\begin{equation}
\vv{\Gamma_{md}}= K_c\vv{\Gamma_d} + 2 K_d \vv{\Gamma_c},
\label{eq_meas2}
\end{equation}
where $K_c = \frac{1} {2} (K_1+K_2)$ and $K_d = \frac{1} {2} (K_1-K_2)$.
Ref. \cite {rodriguescqg1} shows how this equation leads to the requirements on the scale factor matching $K_d$ to $1.5\times 10^{-4}$, with the common-mode acceleration limited to $10^{-12}$\,m\,s$^{-2}$ at the EP frequency about all axes by the DFACS. The DFACS control loop acts over 0.1\,Hz bandwidth, the accelerometer one being about 1\,Hz.
Eq. \ref{eq_meas} can also be adapted when the two test-masses are not perfectly concentric and miscentred by $\vv{\Delta}$, specified to about 20$\,\mu$m along all axes.
\begin{equation}
\vv{\Gamma_d}=\vv{\Gamma_1} - \vv{\Gamma_2}= \delta \vv{g}+\left([\rm{T}]-[\rm{In}]\right)\vv{\Delta},
\label{eq_meas3}
\end{equation}
where $[\rm{T}]$ is the Earth gravity gradient tensor in the instrument's frame, $[\rm{In}]$ is the inertia tensor which is linked to the attitude motion of the satellite defined by: 
$[\rm{In}] = [\Omega] [\Omega]+$$[\dot{\Omega}]$; with $[\Omega]$ the satellite angular velocity tensor.
The ``inertial'' part of the equation depends on the attitude control. With an error allocation of $2\times 10^{-16}$\,m\,s$^{-2}$ for each term, the specifications on the angular motion are :
\begin{itemize}
\item	$[\dot \Omega]$ limited to $10^{-11}$\,rad\,s$^{-2}$ at $f_{\rm{EP}}$ both in inertial and rotating modes;
\item $[\Omega]$ limited to $10^{-9}$\,rad\,s$^{-1}$ at $f_{\rm{EP}}$ in rotating mode.
\end{itemize}
It is worth noticing that a specification of $10^{-9}$\,rad\,s$^{-1}$ at $f_{\rm{EP}}$ (about 3\,mHz in rotating mode) corresponds to an attitude stability of a fraction of $\mu$rad at $f_{\rm{EP}}$. It is obtained thanks to an active attitude control and the use of T-SAGE angular axes for attitude estimation at $f_{\rm{EP}}$.

\subsection {DFACS servo-loop description}

\begin{figure}
\begin{center}
\includegraphics[width=0.8\textwidth]{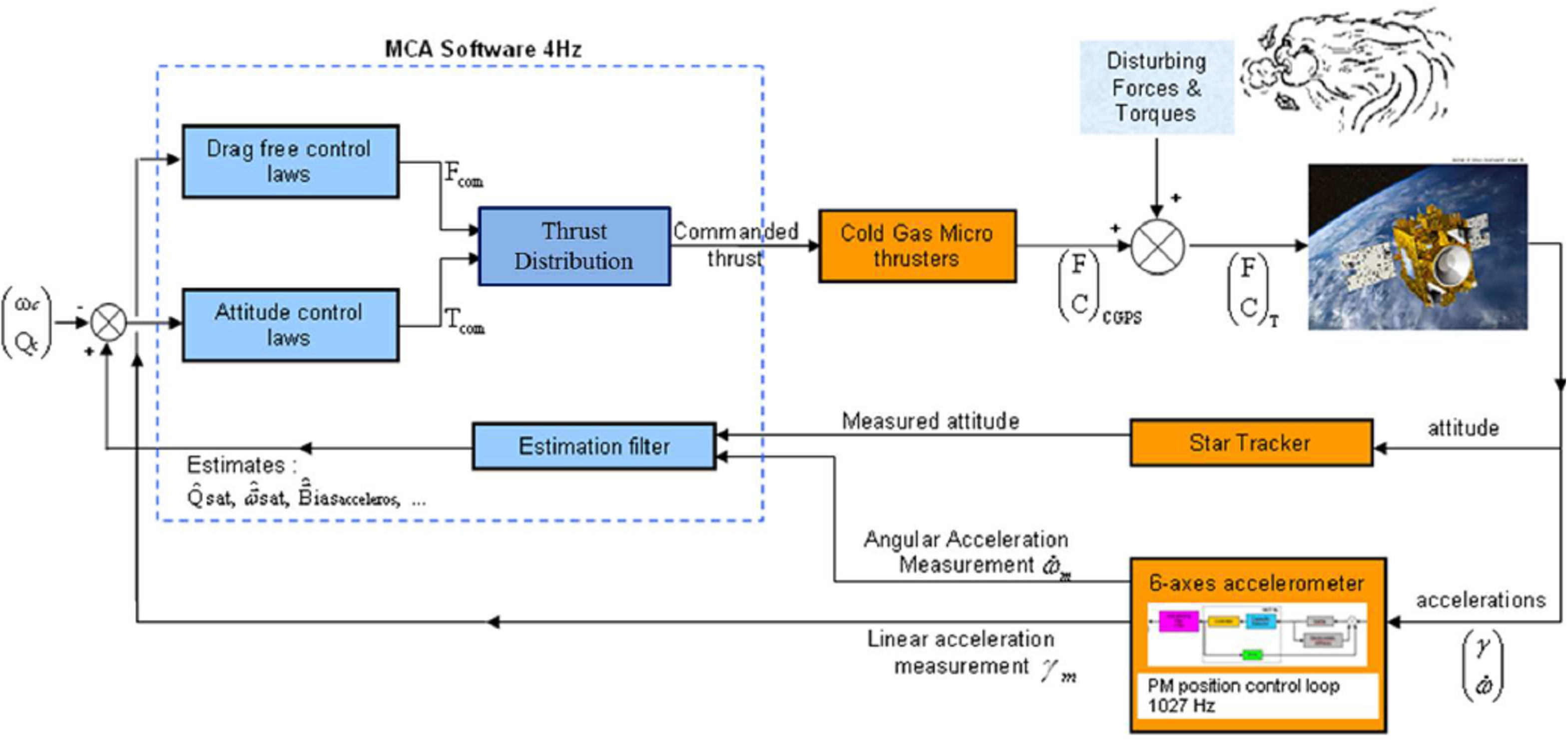}
\caption{DFACS servo-loop. $\binom{\omega_c}{Q_c}$ is the quaternion control setpoint, $\binom{F}{C}_{CGPS}$ is the force and torque applied by the CGPS, $\binom{\gamma}{\dot{\omega}}$ is the applied acceleration to the payload.}
\label{fig_loop}       
\end{center}
\end{figure}

The DFACS control loop (Fig. \ref{fig_loop}) uses the scientific instrument as the main sensor for delivering the linear and angular accelerations and the CGPS as the actuator. The DFACS can be controlled by any combination of the test-masses' measured acceleration. Actually for most of the science sessions one test-mass is selected for the DFACS loop  while few sessions have been performed with the mean value of two test-masses' measurements. The linear acceleration is transmitted almost without processing to the drag-free control laws, the estimated biases being subtracted from the command. The angular acceleration is merged with the star-tracker quaternions with the use of a hybridization Kalman type filter in order to feed the DFACS attitude controller.

Various sets of hybridization filters and controllers are available for different uses. The controller's outputs (force and torque in satellite frame) are projected into thruster axes. As many as 38 loops are closely involved: 6 times 4 loops for the suspension of the test masses, 6 loops for DFACS itself and 8 loops for the local regulation of the thrusters.

\subsection {DFACS software}

\begin{figure}
\begin{center}
\includegraphics[width=0.8\textwidth]{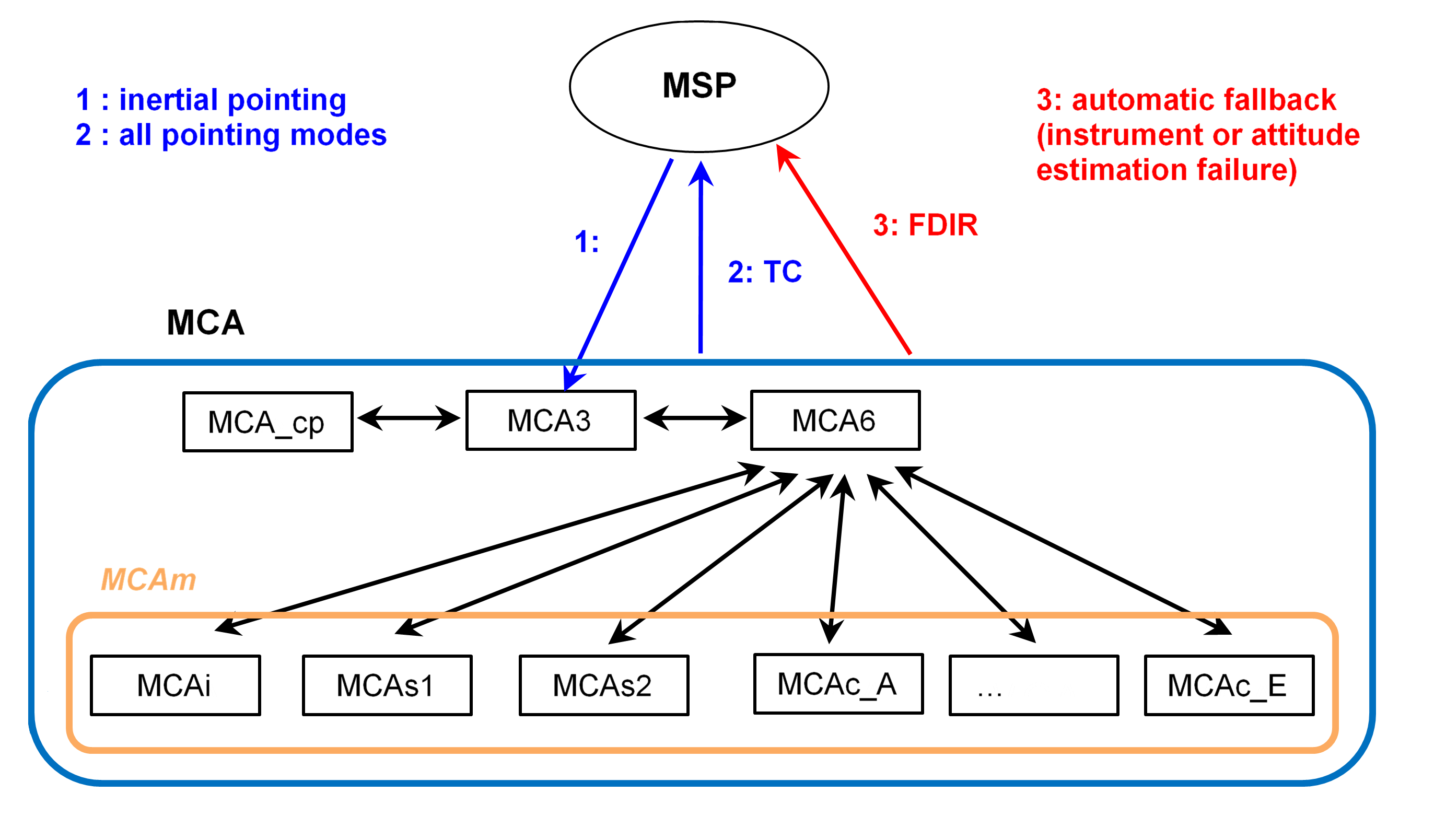}
\caption{DFACS software architecture}
\label{fig_soft}       
\end{center}
\end{figure}

Fig. \ref{fig_soft} presents the software architecture of the DFACS. On top of this  figure, the MSP mode, defined in section \ref{sect_aocs}, performs a fine attitude control through the star tracker measurement and CGPS torques. This mode is also used in collision-avoidance procedures but its main function is to be the gate for the drag-free mode MCA in which one or several test-masses of T-SAGE are used. The MCA mode is made of many tunings: the MCA3 (attitude only) and MCA6 (6-axis control) for instance have low gain robust control and are used to estimate the angular bias of the drag-free test mass, to change the attitude guidance, etc. The MCAcp performs an automatic sequence of tests for the thrusters. The other high gain tunings are dedicated to inertial sessions (MCAi), rotating session velocity 1 (MCAs1), etc.

\begin{figure}
\begin{center}
\includegraphics[width=0.8\textwidth]{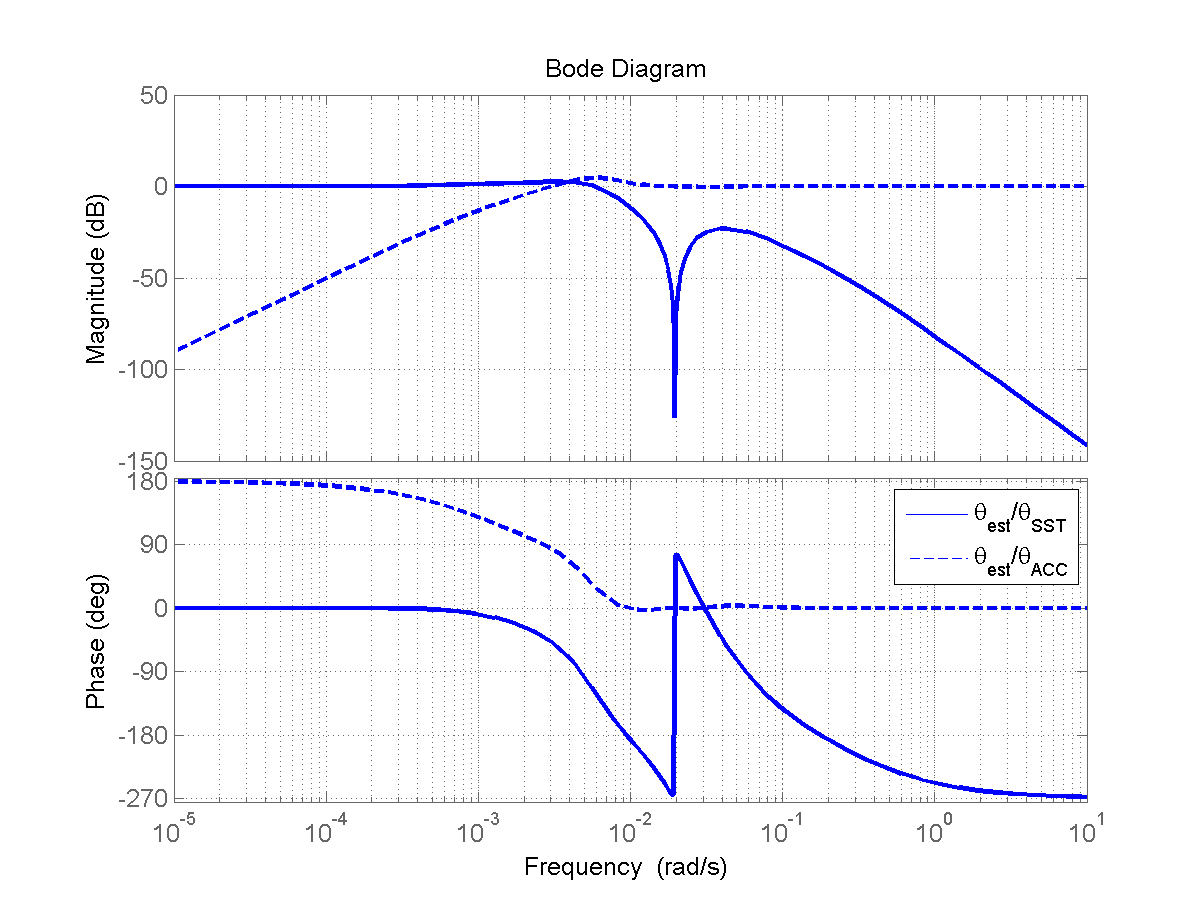}
\caption{Hybridization gain and phase in SpinMax configuration. $\theta_{\rm{est}}$ is the Kalman estimator filter output for the best estimate of attitude, $\theta_{\rm{SST}}$ is the output of the star tracker, $\theta_{\rm{ACC}}$ comes form a double integration of the angular acceleration output of the accelerometer.}
\label{fig_hyb}       
\end{center}
\end{figure}

\begin{figure}
\begin{center}
\includegraphics[width=0.8\textwidth]{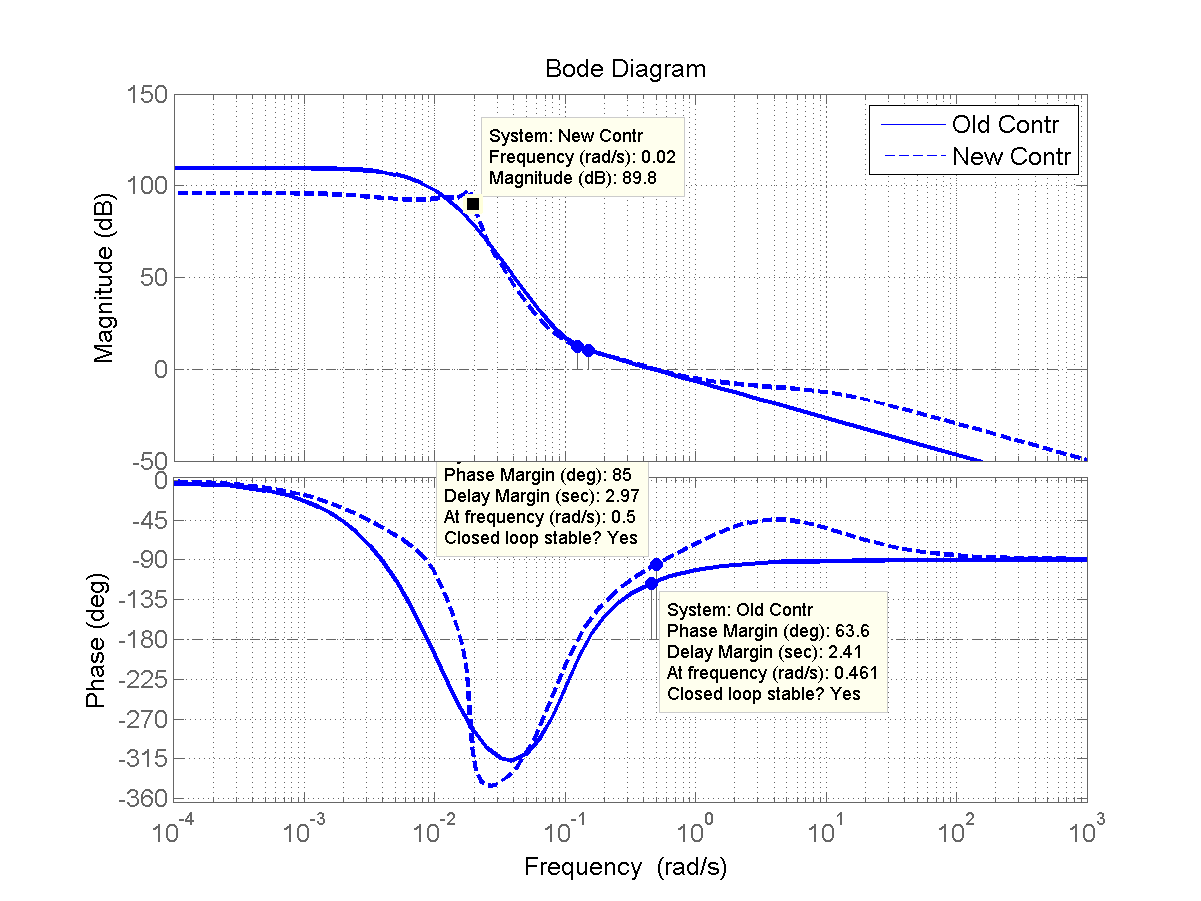}
\caption{Drag-Free controller close loop gain and phase : the continuous line represents the original tuning called old (adapted to spin rate1) and the dotted line is the one called new obtained for SpinMax.}
\label{fig_drag}       
\end{center}
\end{figure}

As shown in Table \ref{tab_freq}, several rate of s/c rotation were tested. Finally the selected EP science sessions were performed with spin rate 2 and 3, the latter called also SpinMax. Fig. \ref{fig_hyb} and \ref{fig_drag} refer to the SpinMax mode. Fig. \ref{fig_hyb} shows the Bode diagram hybridization filter between star tracker quaternions and T-SAGE angular accelerations $[\dot \Omega]$. The hybridization frequency remains high in order to have reasonable convergence duration. The shape of the transfer is modelled using advanced techniques \cite{prieur17, pittet2007}. With this filter, the attitude estimation is not corrupted by errors at $f_{\rm{EP}}$ on the star tracker. 

As air drag acts mainly at $f_{\rm{EP}}$ with 30\,$\mu$N intensity (i.e. $10^{-7}$\,m\,s$^{-2}$ when considering the 300\,kg satellite mass), a minimal rejection of $10^5$ (i.e. 100\,dB) is then needed to reach the required $10^{-12}$\,m\,s$^{-2}$. The challenge becomes to drop the gain above $f_{\rm{EP}}$ quickly enough, to keep stability margins.
Fig. \ref{fig_drag} illustrates the drag-free controller and compares the original tuning to the one actually used in orbit in SpinMax.  Because of lower in flight perturbations, only one tenth of the design hypothesis (air drag $<3\,\mu$N), a 90\,dB rejection was sufficient, but this gain had to be maintained at higher frequency. We managed to have a net delay margin of 1 second at 0.075\,Hz but this control is extremely tight.

\subsection{Attitude guidance in science mode} \label{sect_att}

\begin{figure}
\begin{center}
\includegraphics[width=0.8\textwidth]{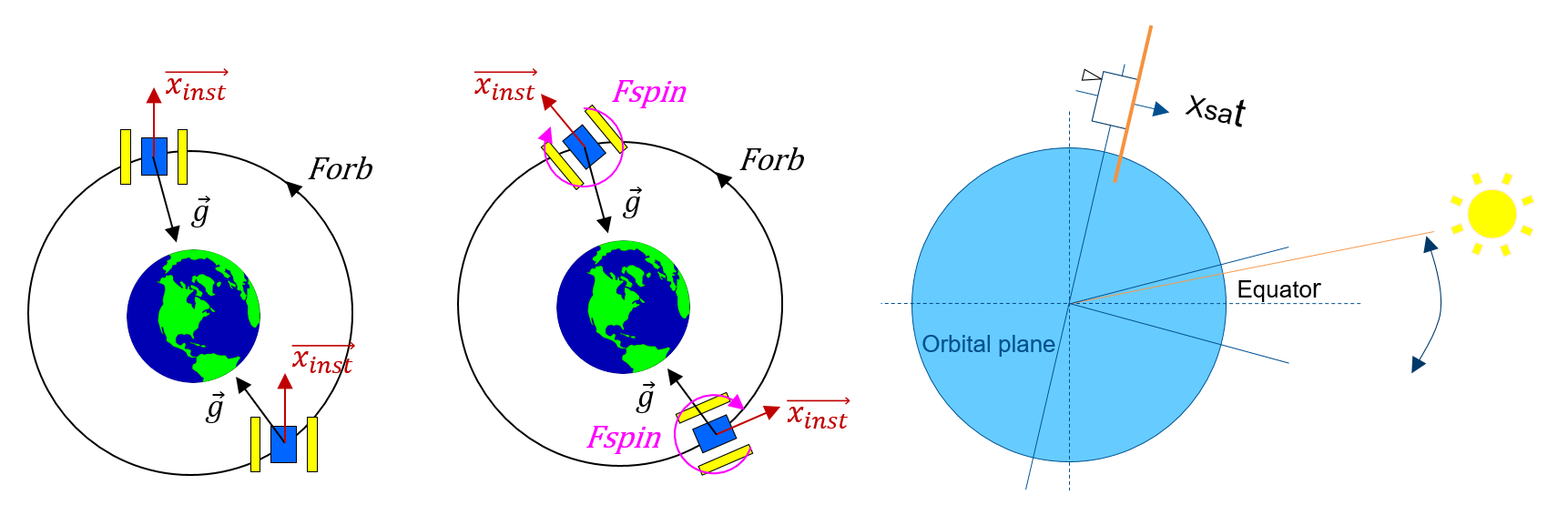}
\caption{Inertial sessions (left), rotating sessions (middle) and side view (right)}
\label{fig_inert}       
\end{center}
\end{figure}

The science measurements were carried out in either inertial sessions or rotating sessions.
During the inertial sessions, the satellite attitude follows the one degree per day drift of the orbital plane: we may use the exact term of quasi inertial mode, but for simplicity in the paper we use the wording inertial mode in oppositon to spin (or rotating) mode. The main axis of the accelerometer ($X_{\rm{inst}} \backsim Z_{\rm{sat}}$) is in the mean orbital plane (Fig. \ref{fig_inert}). 

On the left side of Fig. \ref{fig_inert}, the ``$X_{\rm{sat}}$ to Sun'' angle can be observed between the spin axis $X_{\rm{sat}}$ and the Sun direction. The orbit is entirely sunlit except from May 8$^{th}$ to August 4$^{th}$ where eclipses happen around the southern pole.
This angle plays an important role for thermal and micro-perturbation aspects. Science sessions were interrupted for the eclipse seasons because of non-optimal thermal stability.
In addition, once per month, the satellite has to be slightly tilt to avoid the Moon's glare to the star trackers. During these few days, science sessions are also interrupted. Between two Moon phases, several EP sessions are played surrounded by calibration sessions. Science sessions were defined in the mission design with a typical duration of 8 days \cite{rodriguescqg4}. Because of the higher accelerometer noise in flight than expected in inertial pointing,  rotating sessions with better accelerometer performance were used for the EP test and inertial pointing only used for calibration needing less performance.

As the potential EP violation signal is presumably proportional to the gravitational field \cite {rodriguescqg1}, it is expected to be a sine at the modulation frequency of the gravitational field. So the attitude guidance defines precisesly the EP test frequency. 
During rotating sessions, the satellite is set in rotation around the axis normal to the orbit plane ($Y_{\rm{ins}t}\backsim X_{\rm{sat}}$) at a constant rate $f_{\rm{spin}}$: the gravity signal (i.e. the hypothetical EP signal) is thus modulated at the frequency $f_{\rm{EP}}= f_{\rm{orb}} +f_{\rm{spin}}$. These rotating sessions are designed to last 20 orbits to 120 orbits (about 1 to 8.3 days). With different rotation velocities in order to vary the test conditions, the design was defined on two spin frequencies: $f_{\rm{spin}_1}=\frac{7}{2} f_{\rm{orb}}$ and $f_{\rm{spin}_2}=\frac{9}{2} f_{\rm{orb}}$). After the commissionning phase, a third spin rate was derrived from the two first spin configurations as the SpinMax mode with $f_{\rm{spin}_3}=\frac{35}{2} f_{\rm{orb}}$ (see Table \ref{tab_freq}).

\begin{table}
\caption{\label{tab_freq} Main frequencies of interest.}
\begin{indented}
\item[]\begin{tabular}{@{}llll}
\br
Label & Frequency & Period & Comment \\
   & mHz & sec &\\
\mr
$f_{\rm{orb}}$ & 0.16818 & 5946.026 & Orbital frequency \\
\mr
$f_{\rm{spin}_2}$ & 0.75681 & 1321.339 & Spin frequency 2 \\
\mr
$f_{\rm{spin}_3}$ & 2.94315 & 339.773 & Spin frequency 3  \\
\mr
$f_{\rm{EP}_2}$ & 0.92499 & 1081.096 & EP frequency at s/c spin rate 2  \\
\mr
$f_{\rm{EP}_3}$ & 3.11133 & 321.407 & EP frequency at s/c spin rate 3  \\
\br
\end{tabular}
\end{indented}
\end{table}

The calibration sessions were dedicated to the accelerometer calibration. Based on an inertial pointing, they consist in performing two different types of stimuli:
\begin{itemize}
\item Linear: an additional signal (sinusoidal at $f_{\rm{cal}}=1.22848 \times{}10^{-3}$ Hz) is set on  DFACS accelerometer axis output, leading to a sinusoidal thrust command along the calibration axis. 
\item Angular stimuli: in the same way, the  attitude set-point follows a 50\,mrad amplitude sinusoidal profile at $f_{\rm{cal}}$ and leads to a satellite periodic oscillation.
\end{itemize}

The objective of these stimuli is to generate a reference signal that can be measured by the accelerometer out of the DFACS loop.


\section{Cold Gas Propulsion System - CGPS} \label{sect_cgps}

The European Space Agency (ESA) provided the microthrusters and the CGPS electronics of command.  This contribution was derived from the cold gas propulsion system of the GAIA and the LISA Pathfinder European missions \cite{gaia01}.
CNES was in charge of the implementation of the whole CGPS, including general design, integration and testing. The tanks and the high pressure regulators were off-the-shelf components.

\subsection{CGPS requirements} \label{sect_cgpsreq}

The specification on the CGPS has been established to fullfil the DFACS needs in terms of science performance \cite{rodriguescqg1} and for the satellite maneuver needs. The thrust range command of the CGPS was specified to $500\, \mu$N with a minimum thrust command of $1\, \mu$N and a resolution of 0.2$\, \mu$N. In order to give stability margins on the DFACS servo-loop, the thrust's time response has to be lower than 250\,ms at 63\% of the thrust command. The overshoot (difference between the realized and the commanded impulse) must be lower than 25\% of the commanded impulse on 250\,ms.
In static conditions (constant thrust set point) the thrust noise must be less than 1$\,\mu $N\,Hz$^{-1}$ above 1\,Hz.
The thruster specific impulse dispersion must be lower than 5\% for every thrust set point. The acceleration measured by T-SAGE in variable thrust conditions (for a thrust set point variation of 2$\,\mu$N) has to be lower than $10^{-9}\,$m\,s$^{-2}$.

\subsection{CGPS description} \label{sect_cgpsdes}
\begin{figure}
\begin{center}
\includegraphics[width=0.7\textwidth]{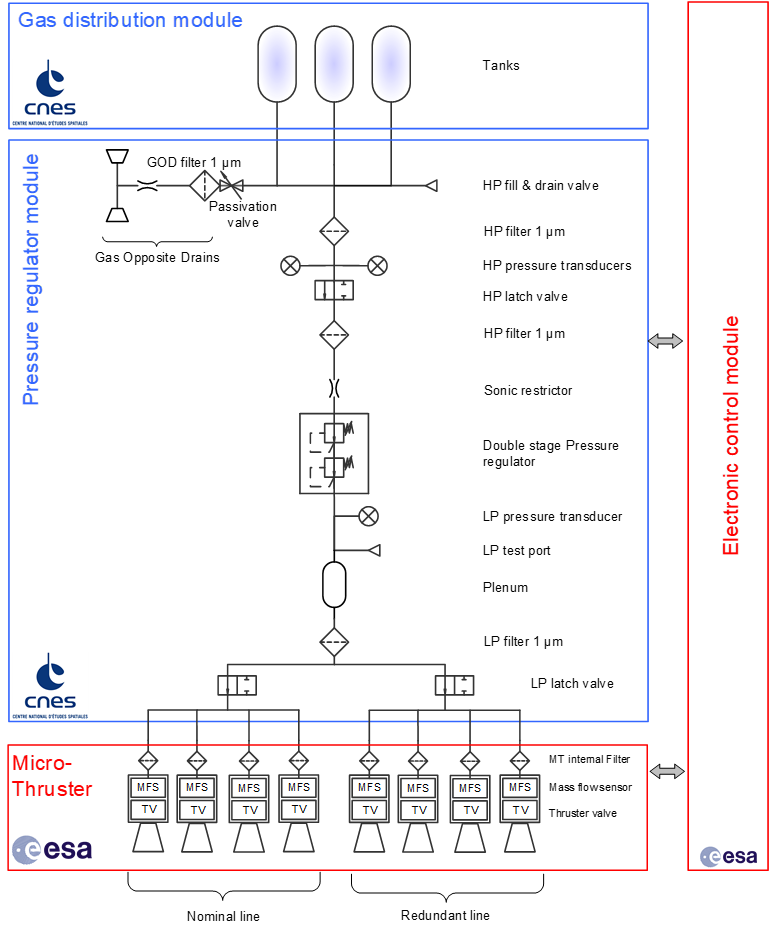}
\caption{CGPSS architecture}
\label{fig_cgps}       
\end{center}
\end{figure}

The CGPS is divided into two identical and independent subsystems called CGPSS (see Fig. \ref{fig_cgps}), accommodated on the two opposite walls of the satellite ($-Z_{\rm{sat}}$ and $+Z_{\rm{sat}}$).

Each CGPSS stores, within 3 connected tanks of 9 liters, a total of 8.25 kg of Nitrogen at 345 bar decreasing to a minimum operational value of 10 bar at the end of the mission. 

From the tanks, a double stage mechanical pressure regulator ensures a regulated 1 bar output pressure to the input of a 0.7 liter plenum. The role of the plenum is to absorb the internal leakage pressure during non-operating phases and to reduce crosstalk between thrusters. The thruster module, fed by the plenum, comprises 4 nominal micro-thrusters and 4 redundant ones. During the mission, the 4 redundant micro-thrusters were fortunately not necessary. Nevertheless, they were tested at the end of the mission for technological survey. For each micro-thruster, a measurement of the gas flow is available through a mass flow sensor (MFS). The flow is controlled continuously by the CGPSS electronics that modify the nozzle cross section of the micro-thruster valve using a piezoceramic actuator. 

The micro-thrusters are accommodated by pairs (the nominal and the redundant one) at the corners of the satellite walls. The position and the orientation of thrusts have been optimized in order to maximize in every direction the total force and torque acting on the satellite.

The CGPSS electronics control module ensures the interface with the on-board computer and runs the flow regulation at 50\,Hz and delivers measurements at 4\,Hz as the payload does: the algorithm includes a specific anti-hysteresis controller in order to improve time responses.

\subsection{CGPS ground tests} \label{sect_cgpstest}
\subsubsection {Microperturbation}

The requirement of a maximum induced acceleration of $10^{-9}\,$m\,s$^{-2}$ has been verified with numerical simulations taking into account the dynamic response of T-SAGE and the main characteristics of the satellite. It has been converted into an equivalent specification of a 1\,kg mass which can move less than 1.2\,nm.
Two sources of microvibration were identified during the preliminary design phase:
\begin{itemize}
\item Moving masses inside mechanism (pressure regulator and micro-thruster). The moving mass requirement on micro-thruster beeing lower than the sensitive level of the microvibration table used for testing, its conformity could not be assessed directly by tests. The moving mass of the plunger and the amplitude of its motion were estimated using a dedicated experiment in order to provide a dynamic model. Using DFACS numerical simulations, the thrust command variation at 4\,Hz was estimated lower than 2$\,\mu$N, leading to a small displacement of the plunger and a maximum value of 0.3\,kg\,nm compliant with the requirement.
\item 	“Clank” created by tank volume variation due to pressure evolution over time. A dynamic model of the pressure regulator was created using the information provided by its  manufacturer; results were compliant with the requirement and no need for further tests was identified. Furthermore, when the tank pressure decreases due to gas consumption or temperature variation, the volume of the gas decreases and a sudden displacement of the tank liner and fiber may occur, creating acceleration spikes. A ground test was performed recording the tank external surface with two high speed cameras, in order to reproduce via a stereoscopic effect a 3D image of the tank while the tank's internal pressure is decreasing. This set-up enables measurement, with a few milliseconds' resolution, of the displacement of the tank external surface and characterization of the sudden variation due to pressure change. Some out-of-specification spikes were observed; however, their frequency was around a few occurrences per hour and their perturbation induced on payload was deemed acceptable.
\end{itemize}

\subsubsection {Fluidics test and thrust regulation}

A simplified engineering model of CGPSS composed of two micro-thrusters, a pressure regulator and an engineering model of the electronic control module was used early in project development to characterize the coupling between CGPS elements. Two major phenomena were observed during these tests.

First, strong oscillations appeared when a thrust step was commanded after a long idle period (i.e. thrust commanded to a fixed low thrust value); the problem was due to the control loop which automatically adjusted thrust hysteresis gain to a too-high value, increasing thruster control loop sensitivity too much and generating oscillations. Once the algorithm was modified and control loop gain values were at limited level, this phenomenon disappeared without degradation of the response time.

Second, a ghost mass flow was observed even when a zero thrust was commanded. This phenomenon is due to the impact of the pressure regulator whose response time is far greater than that of the thruster (several minutes over 250\,ms). The pressure regulator adjusts its working point according to the total gas flow requested by the thrusters. When the global mass flow changes faster than the response time of the pressure regulator, the output pressure never converges to a fixed value. Consequently, a ghost mass flow induced by pressure variation is detected by the MFS. This phenomenon is amplified by the dead volume between MFS and the thrust valve nozzle, which were misestimated during preliminary development.

The discrepancy between the real mass flow and measured value by MFS is roughly 2\% when thrusters are operating. This value has been judged acceptable for the mission thanks to the DFACS closed loop margins.

Nevertheless, the existence of a ghost mass flow changed the in-flight procedure for thruster bias calibration and drive to a modification of the parameter of the leakage detection algorithm implemented on the onboard Fault Detection, Isolation and Recovery (FDIR).

Response time, overshoot and thruster noise requirements involve short-duration phenomena which are not observable on flight due to telemetry sample frequency limitation. For this reason, their validation was performed mainly on the ground as a result of improved observability.

\begin{figure}
\begin{center}
\includegraphics[width=0.6\textwidth]{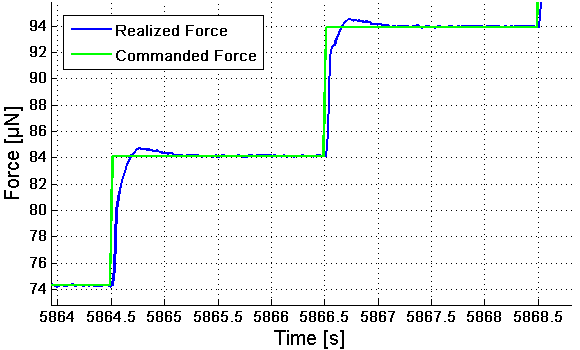}
\includegraphics[width=0.65\textwidth]{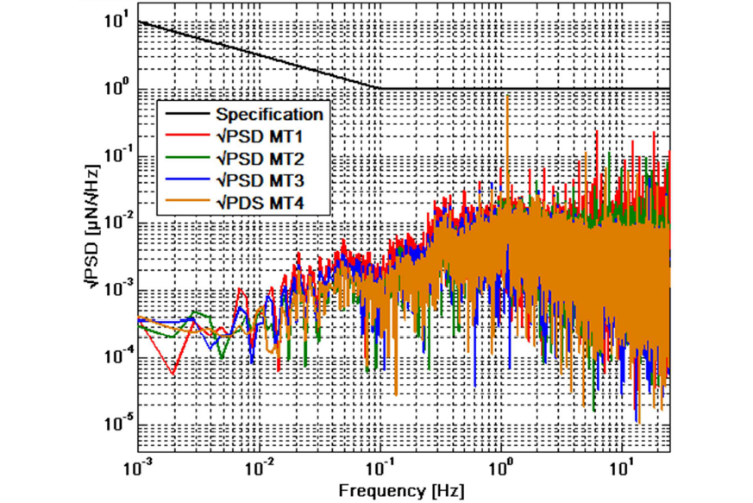}
\caption{Upper: thruster response time and overshoot. Lower: square root of the power spectral density (PSD) of the measured noise for the 4 micro-thrusters compared to the specification.}
\label{fig_cgpsperf}       
\end{center}
\end{figure}

Based on the analysis of several ground test sessions, response time requirements were respected for 97\% of commanded steps and overshoot was respected 90\% of the time. 
The upper pannel of Fig. \ref{fig_cgpsperf} compares the commanded force with respect to the realized one. 

The thrust noise was characterized on ground by two different means: 
\begin{itemize}
\item In the low frequency bandwidth [0.001-0.1] Hz, the noise was directly measured on a micro-balance developped at Onera for the missions GAIA, LISA Pathfinder and MICROSCOPE missions \cite{packan14} ;
\item In the high frequency bandwidth [0.1-10] Hz, the noise was inferrred by the MFS flow measurement recorded at 50Hz frequency sampling with high sensitivity and good resolution.
\end{itemize}

The lower pannel of Fig. \ref{fig_cgpsperf} shows the high frequency noise recorded during satellite thermal vacuum test. Some dispersion between thrusters were observed during ground testing and credited to manufacturing dispersions that lead to a different non-linear response between thrusters (micro-thruster control algorithms use the same parameters for all the thrusters); nevertheless, these differences remain limited and do not affect DFACS. Ref. \cite{lienard17} gives more information about CGPS design.

\subsection{CGPS in-flight performances} \label{sect_perfcgps}

In-flight performance characterizations were limited by T-SAGE's rate acquisition (4\,Hz max). However, the response time of the subsystem is confirmed in space by the excellent behaviour of DFACS, especially in rotating mode which has very small delay margin.

Thrust range, thrust resolution and thruster linearity were directly verified during the in-flight commissionning by comparing the propulsion telemetry with the accelerometer measurements. 
The results fit with the expected values, except for the CGPS scale factor being 10\% under the ground calibration.  The error is attributed to a bias introduced by the mass flow ground equipment used to calibrate the MFS during thruster manufacturing. 
These differences between commanded thrust and performed thrust were indirectly confirmed by gas consumption, which was 10\% less than expected. It was also assessed during tests dedicated to collision-avoidance maneuvers by tracking the satellite position at full thrust range.

Nevertheless, this scale factor does not affect the global behaviour of DFACS because, in close loop, the accelerometer drives its output to null. The error of scale factor is absorbed by the servo loop gain margins. The thruster offsets have been calibrated monthly, showing a drift less than 1.4$\, \mu$N/month on one thruster case, and ten times less for the others.

The gas consumption depends on the type of session \cite{rodriguescqg1, rodriguescqg4} (Table \ref{tab_cgps}). The consumption is determinded by the gyroscopic torque which varies with the square of satellite rotation frequency.
Ref. \cite{lienard17b} gives more information about in-flight CGPS performances.

\begin{table}
\caption{\label{tab_cgps} Total Cold Gas Consumption observed in orbit.}
\begin{indented}
\item[]\begin{tabular}{@{}lll}
\br
Satellite configuration & Mean consumption  & Total consumption \\
  & in g/orbit  &  in g \\
\mr
Inertial  & 1.2 & 2300 \\
\mr
Calibration in inertial pointing & 5.0 & 1400 \\
\mr
Rotation V2 & 1.2 & 1100   \\
\mr
Rotation V3 & 6.6 & 10700  \\
\mr
Transitions & 1.2 & 700 \\
\br
\end{tabular}
\end{indented}
\end{table}


\section{Precise orbit determination} \label{sect_gps}

\subsection{Scientific requirements} \label{sect_gpsreq}

In order to estimate the equivalence principle violation parameter, it is necessary to accurately compute the gravity acceleration and the gravity gradient at each point of the orbit \cite{bergecqg7}. Therefore, a positioning performance of a few meters on the orbit determination is required at the EP test frequency $f_{\rm{EP}}$ and its first harmonics. The requirements on the orbit determination (OD) concern positioning biases (DC errors), and positioning errors at $f_{\rm{EP}}$, 2$f_{\rm{EP}}$ and 3$f_{\rm{EP}}$ frequencies (Table \ref{tab_odreq}). 

The orbit determination is performed with the ZOOMIC automated processing chain, derived from the CNES ZOOM reference software \cite{carrou1}, used for many other missions. A precise orbit was computed for each session of the MICROSCOPE mission within the day following the end of the session. Orbit and associated products (expertise report, error assessment) were then delivered weekly to the scientific mission centre \cite{rodriguescqg4}.
The requirements on orbit determination (OD) have beeen established after several iterations and compromises by considering the mission performance needs \cite{rodriguescqg1} and  the ZOOM performances.

\begin{table}
\caption{\label{tab_odreq} Orbit determination performance requirements on positioning error. In bold: driving errors.}
\begin{indented}
\item[]\begin{tabular}{@{}llll}
\br
Frequency & Radial  & Along-track & Cross-track \\
\mr
DC  & 100\,m & 100\,m & \bf{2\,m} \\
\mr
$f_{\rm{EP}}$ &  \bf{7\,m} &  \bf{14\,m} & 100\,m\\
\mr
2$f_{\rm{EP}}$ & 100\,m & 100\,m & 2\,m   \\
\mr
3$f_{\rm{EP}}$ & 2\,m & 2\,m & 100\,m  \\
\br
\end{tabular}
\end{indented}
\end{table}

However, due to the orbital dynamics, the satellite is mainly sensitive to constant cross-track perturbations and perturbations at the orbital frequency in the orbital plane.
In rotating mode, the orbital error is not significantly affected by the signals at $f_{\rm{EP}}$ or $f_{\rm{spin}}$ (rotation frequency of the satellite).

\subsection{GNSS receiver measurement and processing} \label{sect_sphere}

The receiver delivers position, velocity and time (PVT) and L1 C/A code and carrier phase measurements for 9 channels, with a time to first fix below 90\,s. The default data rate is 10\,s, but a higher rate (2\,s) is possible in technical sessions. The receiver clock drift is about 1\,s/day.

At an altitude of 710km, code and phase observables are affected by ionosphere delays, whose magnitude can reach several tens of meters.

The major advantage of using an ionosphere-free combination is to smartly reject ionosphere-affected measurements in the pre-processing, and then to keep the maximum information without being affected by very disturbing ionosphere effects. The combination used is the semi-sum of code and phase data. The resulting noise is the code noise divided by 2, with an ionosphere effect canceled. 

\subsection{Orbit Determination performance} \label{sect_odperf}

The precise determination of MICROSCOPE orbit relies on the new spatial GPS single-frequency receiver G-SPHERE-S. 
The GPS ionosphere-free-based, PVT-based combined to the One-Way Doppler-based OD are computed for each scientific session, allowing cross-check analysis.
GPS-based OD is the reference orbit. The performance is controlled through several indicators, such as estimated covariance, orbit overlapping analysis, magnitude of estimated parameters and final OD residuals \cite{pascal}.
The OD accuracy estimate for scientific sessions (120-orbit length) is given in Table \ref{tab_odperf}, and is well better than the worst case requirement of $7$\,m. 

\begin{table} [H]
\caption{\label{tab_odperf} Orbit determination estimate accuracy in drag-free mode.}
\begin{indented}
\item[]\begin{tabular}{@{}llll}
\br
Component & Radial  & Along-track & Cross-track \\
\mr
R.M.S OD accuracy  & 10\,cm & 30\,cm & 15\,cm \\
\br
\end{tabular}
\end{indented}
\end{table}


\section{DFACS performances: a worked out example} \label{sect_dperf}

In this section, the  behaviour of DFACS is illustrated by a typical example of one session of 120 orbits. We set our discussion on session 256, a SpinMax session beginning end of April 2017, with the DFACS controlled by  the outer test-mass of SU-EP instrument. 

\subsection{Orbital perturbations}
In inertial and low spin rate sessions, the needed thrust is dominated by magnetic torques to be compensated while in SpinMax and calibration sessions, drag and external forces dominate. In addition, the SpinMax gyroscopic torques due to the angular guidance about the accelerometer placed out of the center of gravity, have a major contribution to the compensation thrust. Finnaly the control is dominated by torques; linear control requires very low thrust as illustrated in Table \ref{tab_force}.

\begin{table}
\caption{\label{tab_force} Session 256, DFACS control force and torque in satellite reference frame.}
\begin{indented}
\item[]\begin{tabular}{@{}lllllll}
\br
& & force     & & \vline  & torque   \\
& & $\mu$N & & \vline  & $\mu$N\,m \\
\mr
& $X_{\rm{sat}}$ & $Y_{\rm{sat}}$ & $Z_{\rm{sat}}$ & \vline \, $X_{\rm{sat}}$ & $Y_{\rm{sat}}$ & $Z_{\rm{sat}}$ \\
\mr
DC & -7.99 & \bf{18.08} & -3.06 & \vline \, -2.91 & \bf{94.82} & \bf{186.38} \\
\mr
$f_{\rm{orb}}$ & 0.38 & 0.03 & 0.02 & \vline \, 0.2 & 0.01 & 0.02 \\
\mr
$f_{\rm{spin3}}$ & 0.23 & 4.90 & 4.94 & \vline \, 1.00 & 4.22 & 4.02 \\
\mr
$f_{\rm{EP}}$ & 0.08 & \bf{2.67} & \bf{2.49} & \vline \, 0.07 & 0.41 & 0.12 \\
\mr
2$f_{\rm{EP}}$ & 0.19 & 1.21 & 0.28 & \vline \, \bf{39.50} & 1.20 & 1.05 \\
\br
\end{tabular}
\end{indented}
\end{table}

In order to save gas, the accelerometer bias is compensated in the loop, and the force at DC is the resulting compensation residual. The air drag acts mainly in the orbital plane ($Y_{\rm{sat}}$, $Z_{\rm{sat}}$) and at $f_{\rm{EP}}$ measured lower than 3\,$\mu$N. When the Sun is distant from $X_{\rm{sat}}$ (by 25deg here), the solar pressure produces a force of 5\,$\mu$N about $Y$ and $Z$ axes at $f_{spin}$ frequency. 
The torques are dominated by gyroscopic effects (static, $Y$ and $Z$ axes) caused by non-diagonal inertia terms ($\Omega_{\rm{spin}}^2$ dependent). The magnetic and gravity gradient torques act at 2$f_{\rm{EP}}$ \cite{hardycqg6} about $X_{\rm{sat}}$.
It is clear that whatever the type of session, the propulsion system is mainly used to control the attitude and very little for performing the drag-free.

\begin{figure}
\begin{center}
\includegraphics[width=0.9\textwidth]{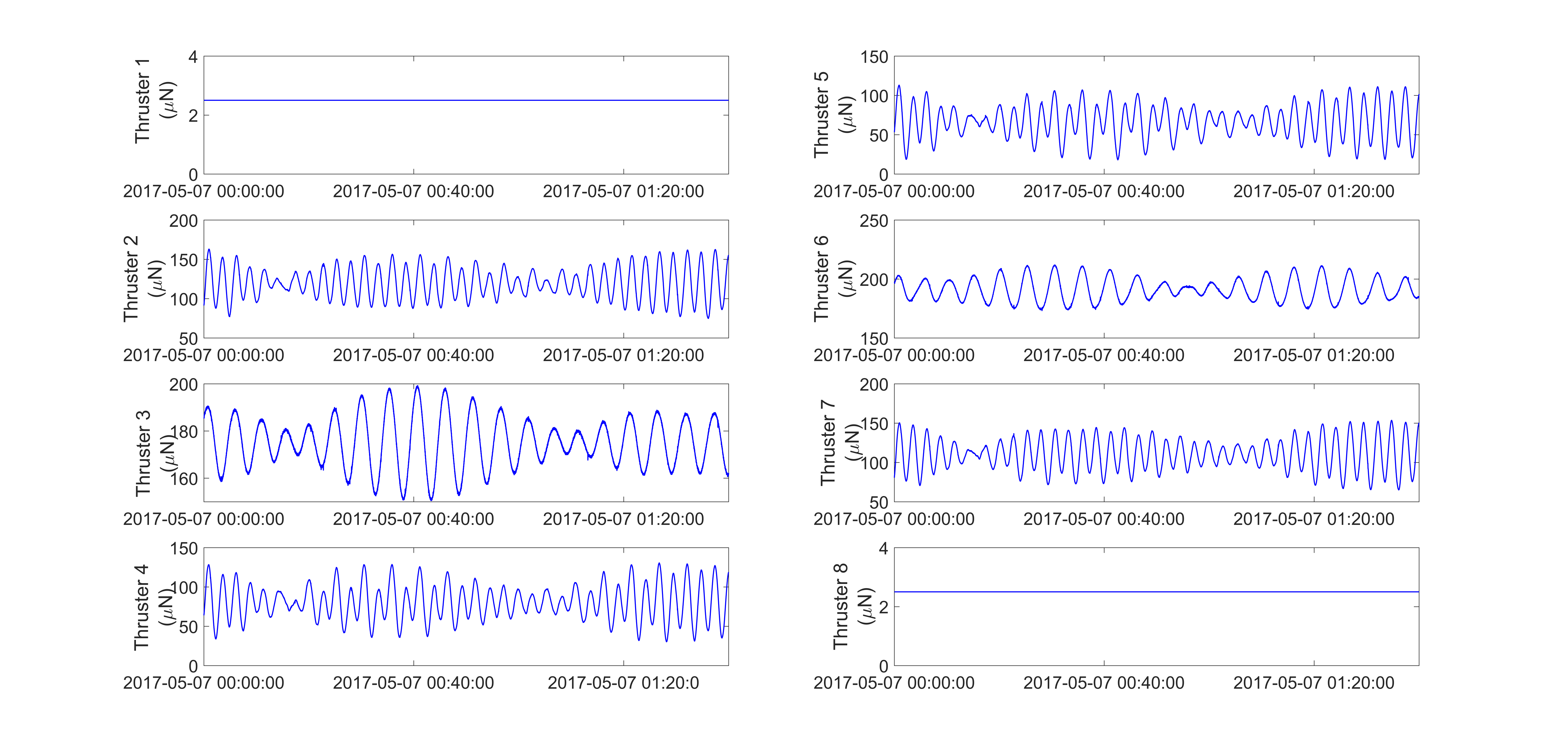}
\caption{Session 256, DFACS commands to propulsion over 1 orbit ($\mu$N)}
\label{fig_prop}       
\end{center}
\end{figure}

Once the command is projected from reference satellite frame to the thruster frame, we observe on Fig. \ref{fig_prop} that the thrusters \#1 placed on the $Z+$ satellite side (upper left curve) and \#8 placed on the $Z-$ satellite side (lower right curve) remain at idle for the entire 8-day session. In contrast, thrusters \#3 and \#6 are loaded close to 200\,$\mu$N. This 4\,Hz plot also shows the very smooth command sent to thrusters in mission modes, ensuring a very low noise environment to the satellite. 

\subsection{Clank perturbations}
\begin{figure}
\begin{center}
\includegraphics[width=0.9\textwidth]{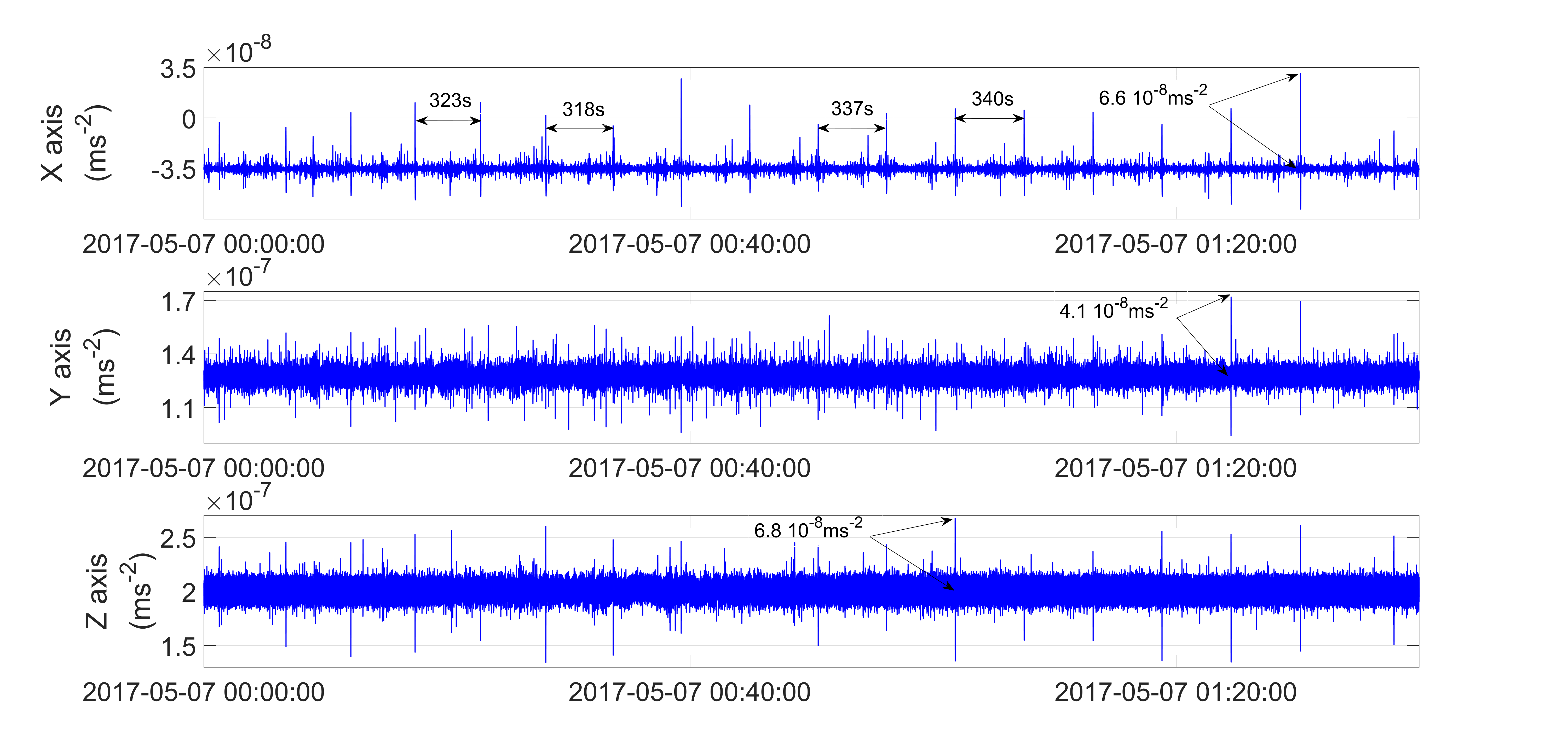}
\includegraphics[width=0.9\textwidth]{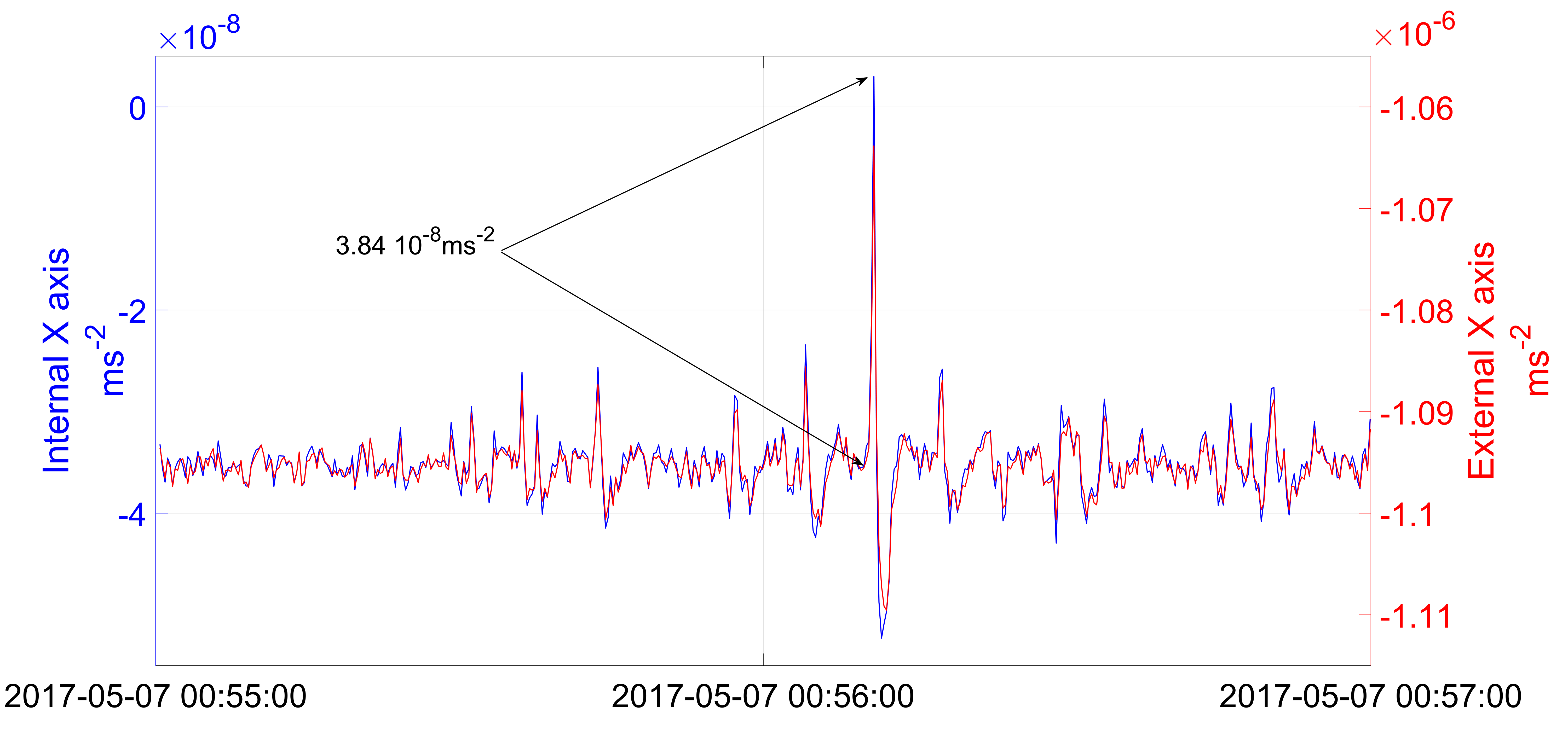}
\caption{Session 256. Upper: linear accelerations measured by SUEP-inner test-mass over 1 orbit along $X_{inst}$, $Y_{inst}$ and $Z_{inst}$ from the top to the bottom ; Lower: acceleration along $X_{inst}$ with a typical spike example (inner test-mass in blue, outer test-mass in red).}
\label{fig_acc}       
\end{center}
\end{figure}

The effect of clanks on the measurement is presented in \cite{bergecqg8} (where they are referred to as "glitches"). We focus here on the satellite design and operation that lead to these clanks.
Fig. \ref{fig_acc} shows an example of linear measurement given by the inner SUEP test-mass over one orbit. We observe spikes almost evenly spread on the orbit with a rate at particular frequencies as $f_{\rm{orb}}$, $f_{\rm{spin}}$ and $f_{\rm{EP}}$. A periodicity of $f_{\rm{spin}}$ ($T_{\rm{spin}}$=340\,sec) would suggest a solar origin and a periodicity of $f_{\rm{EP}}$ ($T_{\rm{EP}}$=321\,sec) would suggest the Earth albedo. Most spikes happen when $–Y_{\rm{sat}}$ panel is oriented toward the Earth, but not systematically. 

This session 256 is one of the most perturbed compared to other SpinMax sessions at other dates. Because of these strong variations with time, spikes may probably have an internal origin (it cannot be Hyper Velocity Impacts coming from the outside, which have a dissymmetrical signature).  Several phenomena can produce these spikes:
\begin{itemize}
\item	Internal clanks due to variation of temperature arround the payload module covered by a MLI, rather at the beginning of the inter-moon period;
\item	Surface clanks due to fluctuation of Sun or Earth illumination of the MLI covering the satellite. The density of spikes could depend on the angular distance between $X_{\rm{sat}}$ (spin axis) and the Sun (see Fig. \ref{fig_inert}).
\end{itemize}
The spikes are lower than $10^{-7}$\,m\,s$^{-2}$, short and with damped sine shape (see Fig. \ref{fig_acc}). Because they are brief with a mean value near 0, their impact on DFACS performance is negligible. Knowing T-SAGE transfer function, we inverse the spike of Fig. \ref{fig_acc} and find that it roughly corresponds to a 100\,g\,$\times \mu$m instantaneous displacement (for example a 2\,g piece of MLI suddenly moving by 50$\,\mu$m). This inversion is sufficiently precise for satellite analysis but not for science correction as we can show in \cite{bergecqg7, metriscqg9} where a particular process was applied to cope with clanks which effect is actually seen differently by the two test-masses.

\subsection{Attitude hybridization}

\begin{figure}
\begin{center}
\includegraphics[width=0.9\textwidth]{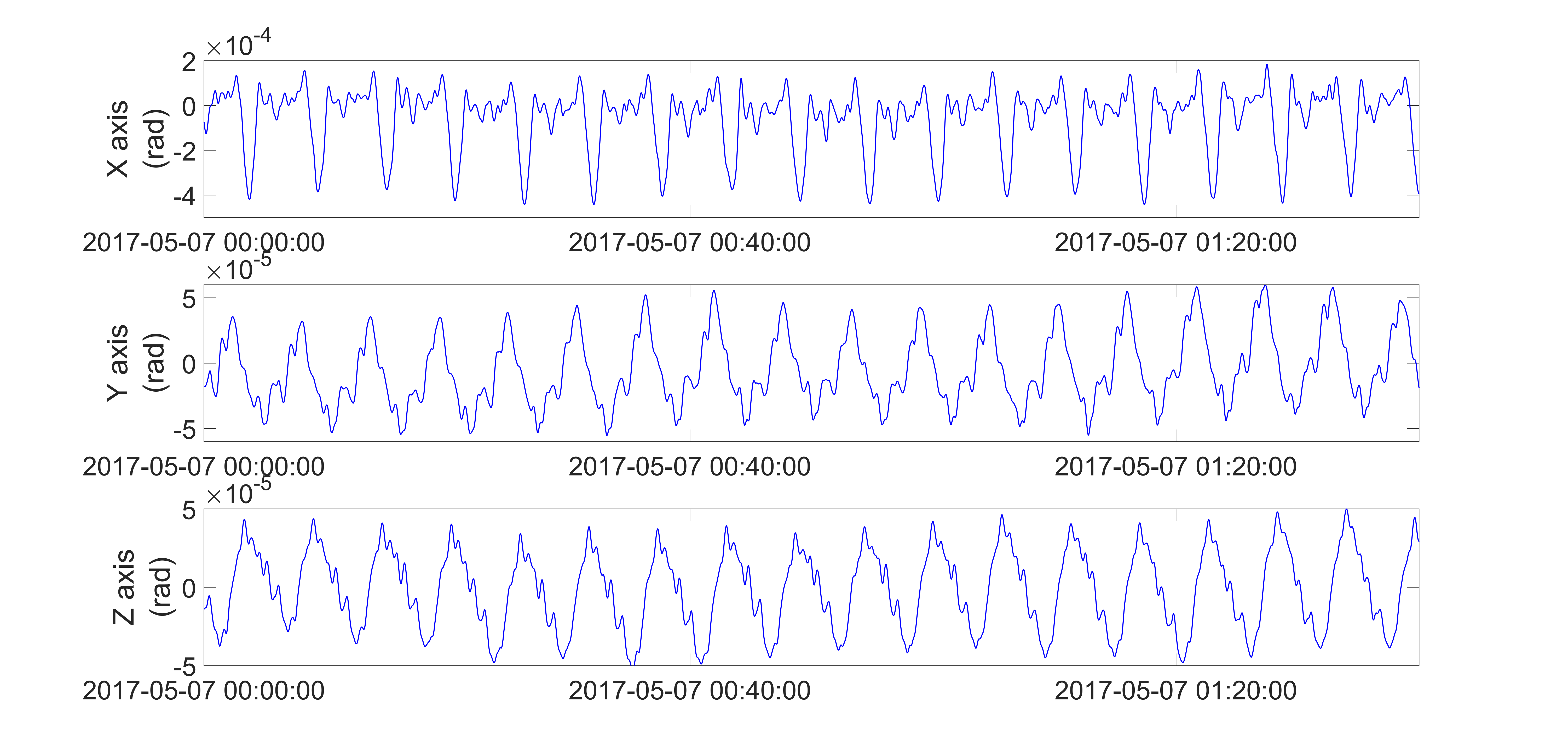}
\caption{Session 256, innovation of the hybridization filter for 1 orbit in radian in the satellite reference frame.}
\label{fig_att}       
\end{center}
\end{figure}

Fig. \ref{fig_att} presents the innovation of the hybridization filter for one orbit (i.e. the gap between the star-tracker measurement and the estimated attitude). We observe a clear periodicity at $f_{\rm{spin}}$ with an amplitude of about 200\,$\mu$rad on $X_{\rm{sat}}$ (spin axis and star-tracker line of sight) and 50$\, \mu$rad on cross axes. This innovation is interpreted as an error from the star-tracker, caused by the rotation of the star pattern in the rectangular field of view. Some days before the beginning of session 256 not in the graph, the amplitude is comparable but the error around $X_{\rm{sat}}$ is mostly at 2$f_{spin}$. This graph shows the interest of the hybridization filter which discards the star-tracker field of view errors; the star-tracker is used only at very low frequency. With the hybridization, the error at $f_{\rm{EP}}$ is about 50 times smaller (3.1\,$\mu$rad at $f_{\rm{EP}}$ about $X_{\rm{sat}}$) than the error of the star-tracker alone. If this measurement was used to control the satellite, we would have a stability of about $6.1\times10^{-8}$\,rad/s, or 61 times above the requirement. The figure  illustrates that the hybridization filter is a key element of performance for DFACS.

\subsection{Drag-Free performance}

\begin{figure}
\begin{center}
\includegraphics[width=0.9\textwidth]{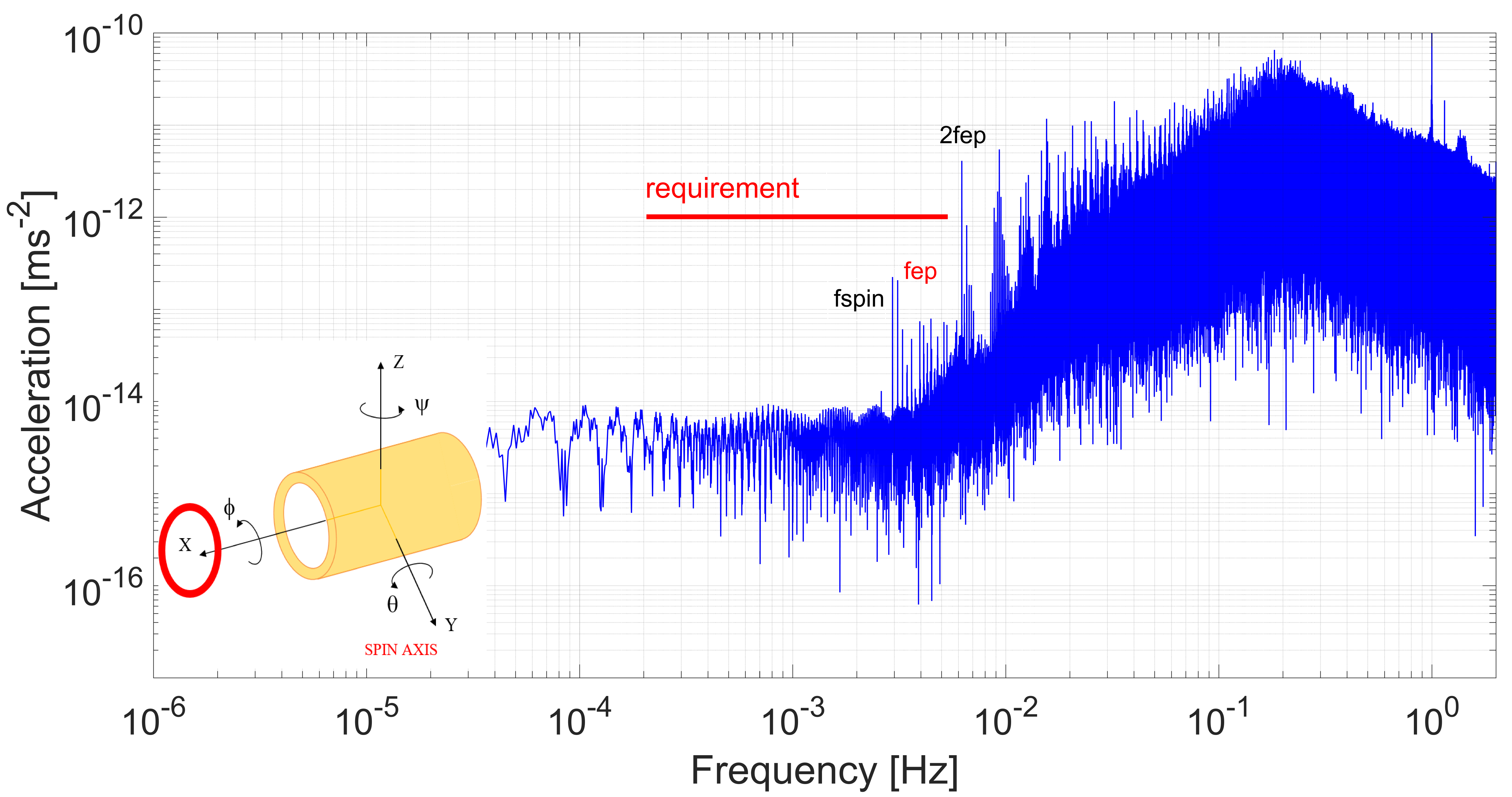}
\caption{Session 256, FFT of the drag-free accelerometer output along $X_{inst}$.}
\label{fig_df}       
\end{center}
\end{figure}

The drag-free performance is estimated through the accelerometer output used for the control loop. Fig. \ref{fig_df} shows the typical residual acceleration observed along $X_{inst}$: $2\times10^{-13}$\,m\,s$^{-2}$ at $f_{\rm{EP}}$. Along $Z_{\rm{inst}}$ the residual acceleration is lower than $2.2\times10^{-13}$\,m\,s$^{-2}$ and lower than $0.1\times10^{-13}$\,m\,s$^{-2}$ along $Y_{\rm{inst}}$.

The control gain quickly drops above $f_{\rm{EP}}$ (3.11\,mHz for the showed session). The 2$f_{\rm{EP}}$ peaks are interpreted as angular to linear coupling due to propulsion: the 39.5$\, \mu$N\,m torque at 2$f_{\rm{EP}}$ is compensated by propulsion which inevitably causes a perturbation in force. This perturbation is rejected by drag-free, but with a limited gain. The first bump at $2\times10^{-2}$\,Hz comes from a transmission of star-tracker stochastic noise, as the second one at $2\times10^{-1}$\,Hz is intrinsically due to the test-mass suspension.  Isolated spectral lines are observed around 1\,Hz and 2\,Hz, caused by aliasing of signal at higher frequencies than the 4\,Hz frequency sampling. 

\subsection{Attitude control performance}

\begin{figure}
\begin{center}
\includegraphics[width=0.9\textwidth]{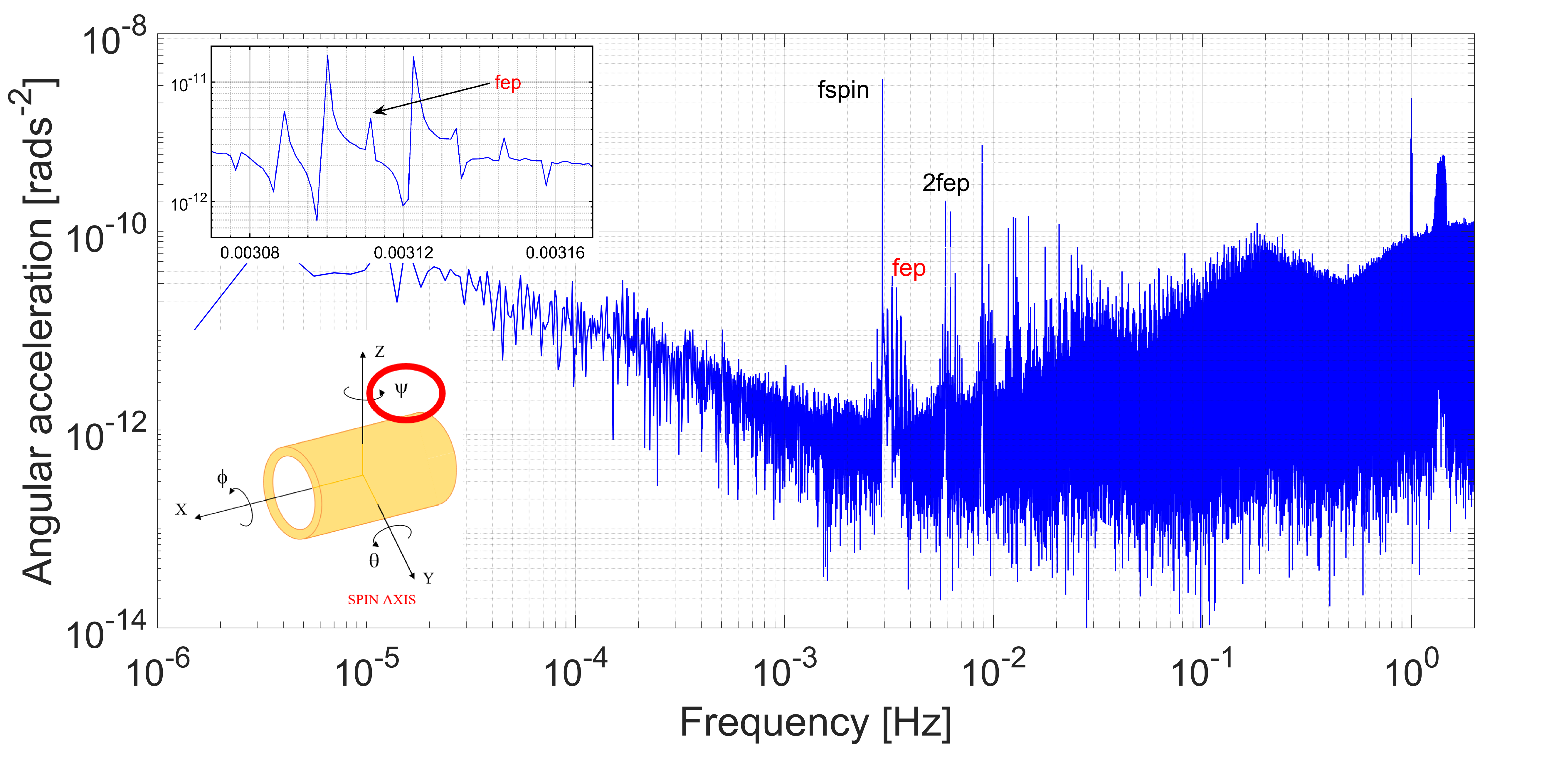}
\caption{Session 256, FFT of the angular accelerometer output about $Z_{\rm{inst}}$ axis, i.e. $\Psi_{\rm{inst}}$.}
\label{fig_psi}       
\end{center}
\end{figure}

Fig. \ref{fig_psi} shows the measured angular acceleration about the $Z_{\rm{inst}}$-axis.  The $f_{\rm{spin}}$ frequency peak is due to the attitude guidance: the spin axis $Y_{\rm{inst}}$ (i.e. $\backsim X_{\rm{sat}}$) follows the orbital plane drift (0.98\,deg/day i.e. $\backsim0.2\,\mu$rad/s) so that the angular acceleration is modulated at $f_{\rm{spin}}$ to $3.7\times 10^{-9}$\,rad\,s$^{-2}$: if the spin axis was strictly inertial, the $X_{\rm{inst}}$ science axis would leave the orbital plane by at least 4\,deg on an 8-day session, breaking small-angle hypothesis used to establish high level requirement tree \cite{rodriguescqg1}.

With 120 orbits, the integration time helps to have a good rejection of the stochastic noise to  $5.64\times 10^{-12}$\,rad\,s$^{-2}$ at $f_{\rm{EP}}$ (Fig. \ref{fig_psi}).
The bunch around 1.5\,Hz and the peak at 1\,Hz are residual frequency aliasing from the sampling process at 4\,Hz. They were reduced with a lower instrument servoloop cut-off frequency and help to limit aliasing of higher frequencies  even if no evidence of an impatc at $f_{\rm{EP}}$ was observed. 

\subsection{Attitude ancillary data performance} \label{sect_oramic}

The scientific exploitation of every session needs the attitude ancillary data \cite{rodriguescqg4}. The CNES ORAMIC tool provides this ``attitude file'' composed of the precise attitude, but also of the angular rate and acceleration. The precise attitude is computed after the day following the end of the session. Attitude and associated products (expertise report, error assessment) are delivered to science mission centre the following week.
The inputs of ORAMIC are the quaternions from the two star tracker camera heads and the angular accelerations from T-SAGE. A particular feature of the algorithms implemented in ORAMIC is that they focus on performance at $f_{\rm{EP}}$ frequency and harmonics. They can be sub-optimal at frequencies of low interest. Each type of satellite modes and thus each type of session has its own algorithm. One star-tracker head or two is selected in association to the considered test-mass for a posteriori restitution.

\begin{figure} [H]
\begin{center}
\includegraphics[width=0.9\textwidth]{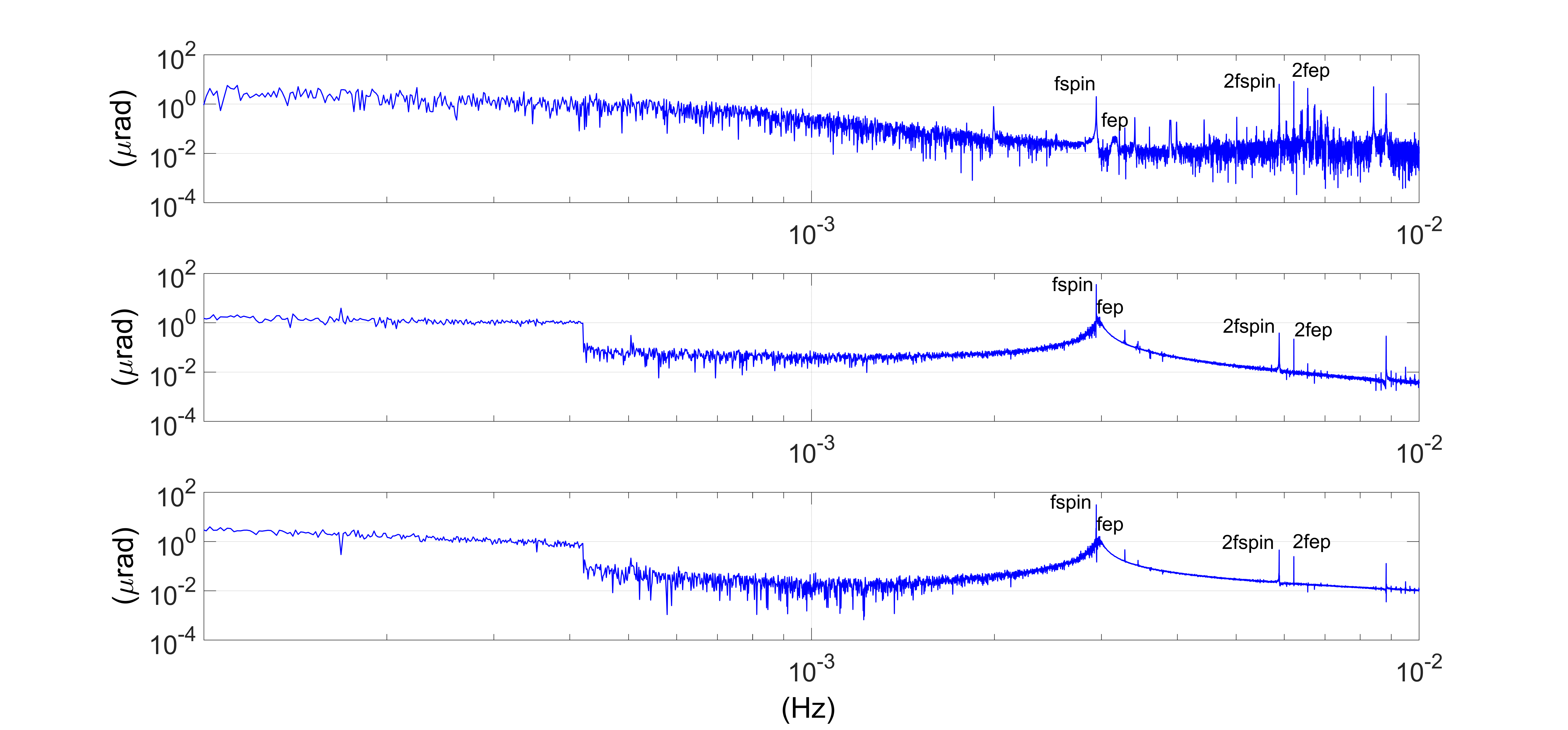}
\caption{Session 256, FFT of attitude restitution in the satellite frame about $X_{\rm{sat}}$, $Y_{\rm{sat}}$ and $Z_{\rm{sat}}$ from top to bottom.}
\label{fig_attr}       
\end{center}
\end{figure}

On one hand, T-SAGE's angular accelerations go through a double integration and on the other hand star-tracker quaternions follow a double derivation process. T-SAGE outputs are kept for most of the spectrum, the star-tracker ones are used at low frequency ($<0.42$\,mHz in the example of session 256) and at particular frequencies like $f_{spin}$ in orbit-plane (unobservable) or at 2$f_{EP}$ (gravity gradient coming from the residuals  non-sphericity of the test mass).
Fig. \ref{fig_attr} shows the FFT of the estimated precise attitude (in $\mu$rad wrt to the guidance profile) for session 256 . The real time attitude estimator on board needs some time to converge (0 to 2 orbits)  because of errors on the T-SAGE estimated biases. In the FFT plot, the sudden change of magnitude is produced by the ``line by line'' frequential hybridization.

\subsection{Summary of DFACS performances for session 256}
Table \ref{tab_dfacp} presents a summary of DFACS performances over the 120-orbit SpinMax session 256, obtained by the ORAMIC tools described in section \ref{sect_oramic}. They comply with the requirements with good margins. The uncertainties are calculated using redundancy of information on angular axes. It can be considered a good session even if the density of spikes is high. 

The gas consumption is 392.9\,g for $+Z_{\rm{sat}}$ (Zp) panel and 392.6\,g for $-Z_{\rm{sat}}$ (Zm) panel (about 3.3\,g/orbit/panel), mainly used to compensate gyroscopic torques.

\begin{table} [H]
\caption{\label{tab_dfacp} Session 256, output of ORAMIC tool for DFACS performance.}
\begin{indented}
\item[]\begin{tabular}{@{}l}

\includegraphics[scale=0.15] {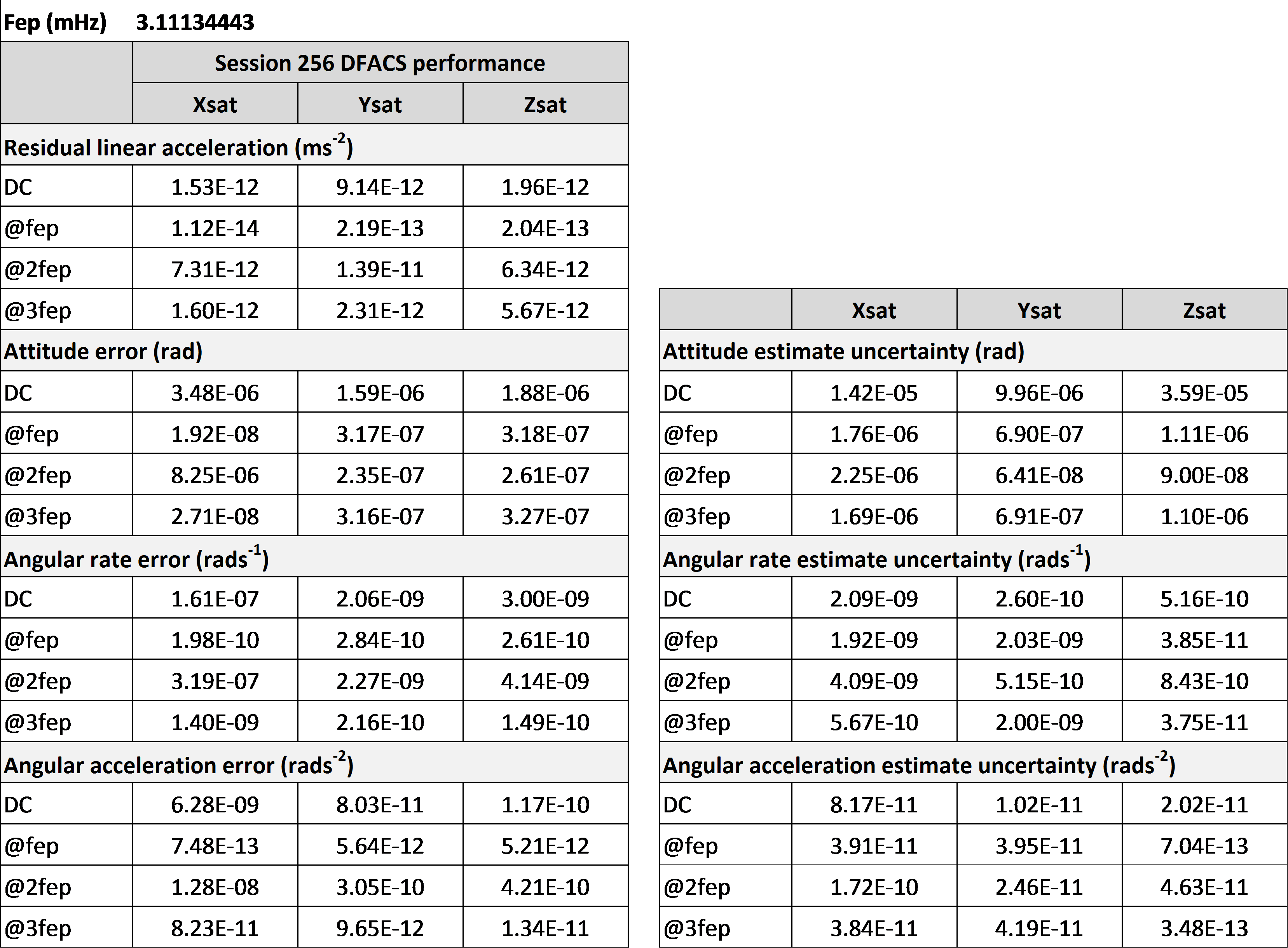} 

\end{tabular}
\end{indented}
\end{table}


\section {Conclusion}

MICROSCOPE satellite flew from April 2016 to October 2018. The performance of the satellite and of the AOCS subsystems were much better than expected. The satellite has been developped to fullfil unprecedent requirements at the level of a micro-satellite line. This was the first time that a drag-free system achieve a pico-g environment ($10^{-12}$ level in linear and angular acceleration SI units) in low Earth orbit. The flexibility to modifiy in ``real time'' the flight configuration helped the science team to use this satellite as a Physics laboratory in space. This largely contributed to the succes of this physics experiment. 

Tools have been developped by CNES to deliver a precise orbit and attitude restitution to the Mission Science Center based at ONERA. These data are used to calculate a precise Earth's gravity field and gradient for the science data process. For each scientific session (calibration or EP test), the performance of the DFACS is also delivered to establish the systematic error budget. To establish the perfromance in terms of extraction of  the E\"otv\"os parameter, it is needed to consider instrument scale factor maching and misalignment. Indeed the performance of the satellite is seen in common mode by the instrument. In ref. \cite{rodriguescqg1, hardycqg6}, the mission performance is thus detailed considering all differential deffects and shows a DFACS error contribution less than a few $10^{-16}$\,m\,s$^{-2}$ to the differential acceleration measurement to be compared to the  $7.9\times10^{-15}$\,m\,s$^{-2}$ mission objective (i.e. $10^{-15}$ on the E\"otv\"os parameter).


\ack

The authors express their gratitude to all the different services involved in the mission partners and in particular CNES, the French space agency in charge of the satellite. This work is based on observations made with the T-SAGE instrument, installed on the CNES-ESA-ONERA-CNRS-OCA-DLR-ZARM MICROSCOPE mission. ONERA authors' work is financially supported by CNES and ONERA fundings.
Authors from OCA, Observatoire de la C\^ote d'Azur, have been supported by OCA, CNRS, the French National Center for Scientific Research, and CNES. ZARM authors' work is supported by the German Space Agency of DLR with funds of the BMWi (FKZ 50 OY 1305) and by the Deutsche Forschungsgemeinschaft DFG (LA 905/12-1). The authors would like to thank the Physikalisch-Technische Bundesanstalt institute in Braunschweig, Germany, for their contribution to the development of the test-masses with funds of CNES and DLR.

\section*{References}
\bibliographystyle{iopart-num}
\bibliography{biblimscope}

\end{document}